\documentclass{mn2e}
\usepackage{graphicx,setspace,pslatex,amssymb}
\usepackage{psfig,subfigure,lscape}
\usepackage{times}
\newcommand{\bgr}{\bibitem[\protect\citename{dummy }1893]{dum}}

\begin{document}
\title[Dust in spiral galaxies]{Dust in spiral galaxies: global properties}

\author[J.\ A.\ Stevens {\it et al.}]
{J.\ A.\ Stevens,${^1}$\footnotemark\ M.\ Amure$^2$ and
W.\ K.\ Gear$^2$ \\ 
$^1$ UK Astronomy Technology Centre, Royal
Observatory, Blackford Hill, Edinburgh, EH9 3HJ \\
$^2$ Department of Physics and Astronomy, Cardiff University,
PO Box 913, Cardiff CF2 3YB1
}


\pagerange{\pageref{firstpage}--\pageref{lastpage}} \pubyear{2005}

\maketitle

\label{firstpage}

\begin{abstract}
We present and analyse high-quality SCUBA 850- and 450-$\mu$m images of 14
local spiral galaxies, including the detection of dust well out into the
extended disk in many cases. We use these data in conjunction with published
far-infrared flux densities from {\em IRAS\/} and {\em ISO\/}, and
millimetre-wave measurements from ground-based facilities to deduce the global
properties of the dust in these galaxies; in particular temperature and
mass. We find that simple two-temperature greybody models of fixed dust
emissivity index $\beta=2$ and with typical temperatures of $25 < T_{\rm warm}
< 40$~K and $10 < T_{\rm cold} < 20$~K provide good fits to the overall
spectral energy distributions. The dust mass in the cold component correlates
with the mass in atomic hydrogen and the mass in the warm component correlates
with the mass in molecular hydrogen. These results thus fit the simple picture
in which the cold dust is heated predominantly by the interstellar radiation
field while the hot dust is heated predominantly by OB stars in more active
regions, although we argue that there is some mixing. The mean gas-to-dust mass
ratio is $120\pm60$, very similar to that found within our own galaxy and
roughly a factor of 10 lower than that derived from {\em IRAS\/} data
alone. The gas-to-dust mass ratios in the warm, molecular component are on
average higher than those in the cold, atomic component. We compare our
modelling results with similar results for more luminous spiral galaxies
selected at far-infrared wavelengths by the SCUBA Local Universe Galaxy Survey
and find that whilst the total dust mass distributions of the two samples are
indistinguishable, they have significantly different dust temperature
distributions in both the warm and cold components. We suggest that this
difference might be related to the level of star-formation activity in these
systems, with the more active galaxies having more intense interstellar
radiation fields and higher dust temperatures.
\end{abstract}


\footnotetext{E-mail: jas@roe.ac.uk}

\section{Introduction}

The interstellar media (ISM) of galaxies are both the archaeological record of
metal production during the star-formation history of the galaxy and the
reservoir from which current star-formation draws its material. Despite the
great interest in tracing the star formation history and evolution of galaxies
surprisingly little is known about the largest sink for metals in the ISM,
namely dust grains (see e.g. Whittet 1992; Edmunds 2001).

After the {\em IRAS\/} all-sky survey many studies were published of the dust
properties in both `active' and `normal' galaxies. An influential study was
conducted by Devereux \& Young (1990; henceforth DY90) in which they derived
dust masses from {\em IRAS\/} observations of a sample of 58 spiral galaxies
and by comparison with CO (as a tracer of H$_2$) and H~{\sc i} measurements of
the gas concluded that the average derived total gas-to-dust mass ratio was
$\simeq$1100, compared to the value $\simeq$100 generally found in our own
galaxy. DY90 and others recognized that this almost certainly indicated that
{\em IRAS\/} measurements at 60 to 100~$\mu$m were simply not sensitive to the
cool dust that may dominate the total dust mass of a galaxy.

Following the launch of {\em ISO\/} and the commissioning of SCUBA on the JCMT
and MAMBO on the IRAM 30-m telescope, longer wavelength measurements of dust in
spiral galaxies have become possible and unsurprisingly several detailed
studies have confirmed that indeed there is more dust in spiral galaxies than
{\em IRAS} data alone would suggest (Gu${\rm \acute{e}}$lin et al. 1993; Braine
et al. 1995, 1997; Neininger \& Dumke 1999; Neininger et al. 1996; Dumke et
al. 1997; Alton et al. 1998; Haas et al. 1998; Davies et al. 1999; Contursi et
al 2001; Popescu et al. 2002; Hippelein et al. 2003; Meijerink et al. 2004).
The SCUBA Local Universe Galaxy Survey (SLUGS; Dunne et al. 2000; Dunne \&
Eales 2001) made brief observations of a large sample of {\em IRAS}-selected
galaxies. They concluded that two dust temperature components were required to
fit the spectra, both with an emissivity index $\beta \simeq 2$. In an
overlapping study, James et al. (2002) used SCUBA observations of galaxies with
known metallicity and the assumption that a constant fraction of metals in a
galaxy are bound up in dust grains to derive an absolute dust emissivity
consistent with the an extrapolation of the far-infrared results of Hildebrand
(1983) and Casey (1991) with $\beta \simeq$2.

We have also previously published a study of the dust in NGC~3079 (Stevens \&
Gear 2000). This paper is a follow-up with a much larger sample. For reasons of
brevity we do not in this paper attempt a very detailed study of the morphology
of each individual galaxy, which the quality of the images certainly merit, but
rather concentrate on the more global properties of the sample and defer more
detailed comparison with e.g. CO and H~{\sc i} images to a future paper.

\begin{table*}
\begin{minipage}{125mm}

\caption{\ Parameters of the spiral galaxies in this study}
\label{table:params}
\begin{tabular}{lccccccc}
\hline
(1)&(2)&(3)&(4)&(5)&(6)&(7)&(8)\\
{Source}&{Distance}&{D$_{\rm 25}$}&{D$_{\rm 25}$}&{Type}&{Chop PA}&{Data source}&{Environment}\\
& {(Mpc)} & {(arcmin)} & {(kpc)} & & ($^{\circ}$) & &\\\hline
NGC 157&   35.0 ~~~& 4.27 ~~~& 43.4 ~&SAB(rs)bc   &$90$  &Archive&Field  \\
NGC 660&   19.6 ~~~& 9.12 ~~~& 52.0 ~&SB(s)a pec  &$80$  &Archive&Field  \\
NGC 1808&  17.1$^a$ ~~~& 7.24 ~~~& 36.0 ~&SAB(s)a &$43$  &Archive&Field  \\
NGC 2903&    9.3 ~~~&12.59 ~~~& 34.2 ~&SAB(rs)bc   &$90$ &Author &Field  \\
NGC 3310&   21.3 ~~~& 3.63 ~~~& 22.4 ~&SAB(r)bc pec&$90$ &Archive&Field  \\
NGC 3368&   15.5 ~~~& 7.08 ~~~& 31.8 ~&SAB(rs)ab   &$90$ &Author &Group  \\
NGC 3628&    6.7 ~~~&14.79 ~~~& 28.8 ~&Sb sp pec   &$14$ &Archive&Triplet\\
NGC 4303&   20.0 ~~~& 6.03 ~~~& 35.1 ~&SAB(rs)bc   &$90$ &Author &Virgo  \\
NGC 4388&   20.0 ~~~& 5.13 ~~~& 29.8 ~&SA(s)b sp   &$0$  &Archive&Virgo  \\
NGC 4402&   20.0 ~~~& 4.07 ~~~& 23.7 ~&Sb sp       &$0$  &Author &Virgo  \\
NGC 4414&   14.4 ~~~& 3.63 ~~~& 15.2 ~&SA(rs)c     &$65$ &Archive&Field  \\
NGC 4501&   20.0 ~~~& 6.92 ~~~& 40.3 ~&SA(rs)b     &$50$ &Author &Virgo  \\
NGC 4631&   12.8 ~~~&15.14 ~~~& 56.2 ~&SB(s)d sp   &$0$  &Author &Pair   \\
NGC 5907&   15.6 ~~~&12.30 ~~~& 55.8 ~&SA(s)c sp   &$65$ &Archive&Group  \\\hline
\end{tabular}

(1) Source name.
(2) Distance to the galaxy calculated from the Hubble Law after
applying a solar motion correction and using $H_0 = 50~\rm{km~s}^{-1}
\rm{Mpc}^{-1}$ (Young et~al. 1995).  Virgo cluster members are assumed to be at
a distance of 20 Mpc. $^a$The distance estimate of NGC~1808 is taken from Sofue
(1996) using $H_0 = 50~\rm{km~s}^{-1} \rm{Mpc}^{-1}$.
(3) Optical angular diameter out to the 25$^{\rm th}$ mag
arcsec$^{\rm -2}$ isophote from de Vaucouleurs, de Vaucouleurs \& Corwin
(1976).
(4) Linear diameter calculated from columns (2) and (3).
(5) Morphological type from de Vaucouleurs et al. (1976).
(6) Chop position angle measured east of north in the RA-Dec
co-ordinate frame.
(7) Details of whether the galaxy was observed by the Author or
taken from the SCUBA archive.
(8) Environment in which the galaxy resides.

\end{minipage}
\end{table*}

\section{Sample}

The galaxies observed in no sense constitute a complete or well-defined
sample. They were chosen on the basis of being (a) large enough for SCUBA to
resolve the core, disc, spiral arm and bar (if present) regions but (b) small
enough that it was feasible to chop off the disc (c) bright enough at
far-infrared wavelengths to allow mapping with SCUBA with reasonable
integration times (of order a few hours) and (d) with published H~{\sc i} and
CO measurements. Only spiral galaxies were chosen to enable a comparison of gas
and dust content with the Milky Way. Finally, they had to be observable during
the time slots awarded, or have data available in the archive. The final sample
of 14 sources is given in Table~\ref{table:params}. The 850-$\mu$m data for
NGC~5907 have been published by Alton et al. (2004).

\section{Observations and Data Reduction}

The Galaxies marked `Author' were observed by the authors over the period
February 1998 to January 2001. Those marked `Archive' were observed by others
over roughly the same period and the data obtained from the JCMT archive and
reduced by the authors.

All observations were obtained using SCUBA in `jiggle-map' mode with a 120
arcsec chop throw; chop position angle are listed in Table~\ref{table:params}.
In jiggle-mode the array is shifted sequentially through 64 positions offset by
$\sim3$ arcsec in order to fully-sample the sky at both 450 and 850~$\mu$m. One
second is spent in each position, and the telescope is 'nodded' every 16
positions.  In many cases, and particularly for the more edge-on sources,
several jiggle-maps slightly offset from each other were taken and then
combined together, weighted by the rms noise in each individual map, to produce
the final image. Skydips were used to monitor the atmospheric emission and flux
calibration was performed against Mars, Uranus, CRL~618 and CRL~2688 with at
least one and usually two of these sources being observed on each night that
data were taken. The flux-conversion factors (FCF) derived from these
calibrations were all in good agreement with standard values published on the
SCUBA website. We estimate the total uncertainty in the flux densities to be
approximately 25 per cent at 450 and 20 per cent at 850~$\mu$m respectively,
dominated by the uncertainty in determining the atmospheric transmission
correction. The SCUBA filters are very narrow, making colour-correction
unnecessary.

Data were reduced with the standard STARLINK {\sc surf} package (Jenness et
al. 2002) with the addition of (1) a custom designed frame-by-frame baseline
removal algorithm and (2) modified versions of the {\sc setbolwt} and {\sc
rebin} tasks. Since these additions were written by ourselves we elaborate on
them below.

\begin{figure}
\setlength{\unitlength}{1in}
\begin{picture}(3.0,4.5)
\includegraphics{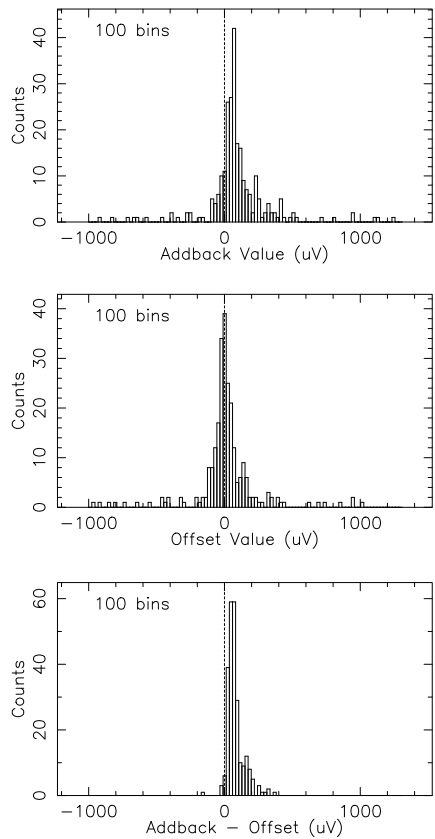}
\end{picture}
\caption[dum]{The distribution of `addback' values that were first removed and then added
back onto the data by the {\sc surf} task {\sc remsky} ({\em top}).  The distribution
of baseline offsets on the individual re-gridded images measured using the
STARLINK graphical analysis tool {\sc gaia} ({\em middle}).  The difference
between the addback value and baseline offset for each image ({\em bottom}).
The units are $10^{-6}$ Volts.}
\label{fig:red}
\end{figure}

\subsection{Sky removal}

The first two steps in the reduction correct for the beamswitching and
atmospheric extinction. However, some residual varying sky emission always
remains, this emission is strongly correlated between pixels across the array
and can be decreased by subtracting the median signal measured in one or
several source-free or `sky' pixels from all the others. Unfortunately, because
the galaxies are so extended, it was not possible in many cases to find genuine
'sky' pixels, however since the source flux is also extended and faint it is
not strongly detected above the sky in any 1-second jiggle point and so the
median value of {\it all} pixels in the array provides a good measure of the
sky emission to be subtracted.

It was found that in some cases the sky emission was clearly not averaging to
zero, and hence there was a danger of adding or subtracting spurious flux
density from the image. To cure this problem, carefully-selected regions devoid
of source emission were identified on the images and defined to be zero to
remove the underlying sky emission baseline. It was easier to identify such
emission-free regions in the final map than to identify 'sky' bolometers
because the jiggling action and sky rotation during an observation means that
each bolometer actually moves over a significantly larger area of the
source. The regions chosen were also double-checked against other wavelength
images from various archives and confirmed as being unlikely to have
significant underlying emission.

The distribution of addback values, of non-zero integrated sky emission
baseline-offsets and their 'differences' are shown in Fig.~\ref{fig:red}. A
total of 124 maps were reduced at both 850 and 450~$\mu$m giving a total of 248
values.

Since the addback value is the average of the sum of the source flux density
seen in the median bolometer value and the sky emission for each data point,
the difference is the source flux density seen. This is roughly constant due to
the source selection process, i.e. large galaxies which fill the SCUBA field
with surface brightness large enough to be detected in a few hours. Almost all
values are reassuringly positive, the four negative examples are from noisy
maps. The positive tail is due to the relatively brighter galaxy cores.
Further discussion of various methods of sky-noise removal and problems
associated with them can be found in Amure (2003).

\subsection{Regridding}

After sky removal each observation was calibrated with the appropriate FCF and
the data were processed with the {\sc surf} task {\sc despike} which
sigma-clips after regridding onto sky coordinates.  We then used adapted
versions of the tasks {\sc setbolwt} and {\sc rebin} to weight and regrid the
data onto an RA/Dec coordinate frame. The tweaked {\sc setbolwt} task was first
used to calculate the standard deviation of the data in each bolometer for the
whole dataset to be regridded. We adapted the `median' option in {\sc rebin} so
that it uses these standard deviations to calculate a weighted mean and
weighted noise map. For some of the galaxies several jiggle-maps were added
together, after determining a `baseline' value for each map from the mean of
six `source-free' regions, setting these all equal to zero, then averaging
pixel values in the overlap regions weighted by the inverse-square of the mean
noise level in the `source-free' regions in each individual map. The maps were
made with 2-arcsec pixels and then convolved with a Gaussian to improve the
signal-to-noise (S/N).  We found that a 3-arcsec Gaussian increased the S/N
sufficiently on all 850-$\mu$m maps to produce a 3-$\sigma$ detection across
most of each galaxy without degrading the resolution significantly. The final
resolution of these maps is $\sim15$~arcsec. The high-quality 450-$\mu$m maps
were smoothed with a 2 arcsec Gaussian giving resolutions of $\sim8$
arcsec. However, several 450-$\mu$m maps had to be smoothed with 4 arcsec
Gaussians giving resolutions of $\sim9$ arcsec -- these galaxies are NGC~3310,
NGC~3368, NGC~4303, NGC~4402, NGC~4501 and NGC~5907.

This procedure provides a number of advantages over the default {\sc
surf}. Primarily, since we have a noise map, we know the S/N at each position
on the image and do not have to correct statistically for correlated noise when
estimating noise levels from smoothed signal maps (see Dunne \& Eales 2001). We
can then, for example, plot meaningful S/N contours on the signal maps rather
than applying noise estimates calculated from signal-free regions of the signal
image to regions where the noise is different.

\begin{figure*}
\setlength{\unitlength}{1in}
\begin{picture}(7.0,3.7)
\includegraphics{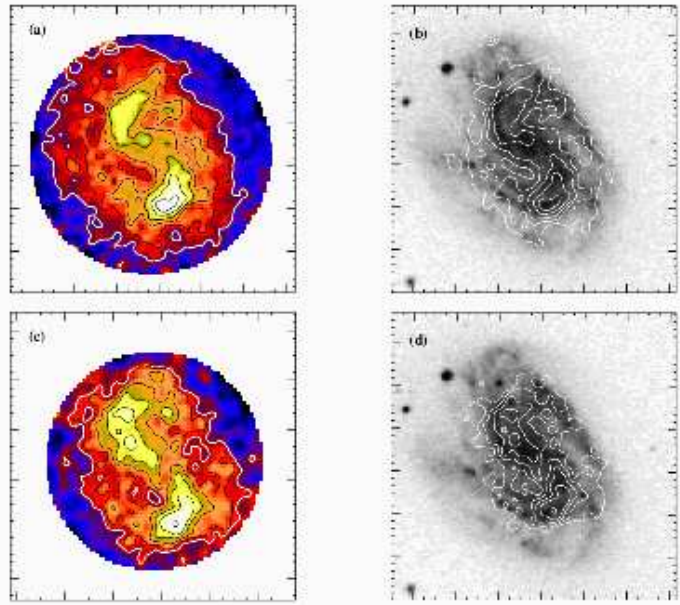}
\end{picture}
\caption[dum]{Images of NGC 157 (a) 850-$\mu$m image as contoured  colour scale. The
white contour is the 3-$\sigma$ level from the signal-to-noise image. Black
contours show flux densities of 15, 21, 27, 36 and 45~mJy\,beam$^{-1}$. (b) DSS
image overlaid with the 850-$\mu$m contours. (c) 450-$\mu$m image as contoured
colour scale. The white contour is the 3-$\sigma$ level from the signal-to-noise 
image. Black contours show flux densities of 60, 80, 100, 120 and
160~mJy\,beam$^{-1}$.  Panels (a) and (b) are $195$ arcsec square while panels 
(c) and (d) are $170$ arcsec square.}
\label{fig:ngc157}
\end{figure*}

\begin{figure*}
\setlength{\unitlength}{1in}
\begin{picture}(7.0,3.8)
\includegraphics{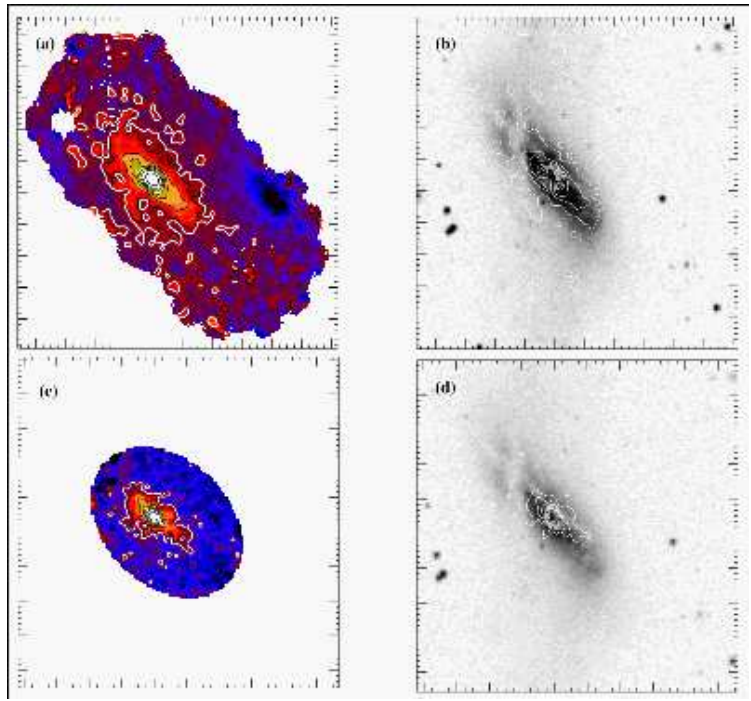}
\end{picture}
\caption[dum]{Images of NGC 660 (a) 850-$\mu$m image as contoured colour scale. The
white contour is the 3-$\sigma$ level from the signal-to-noise image. Black
contours show flux densities of 20, 50, 80, 110, 140, 170, 200 and
230~mJy\,beam$^{-1}$. (b) DSS
image overlaid with the 850-$\mu$m contours. (c) 450-$\mu$m image as contoured
colour scale -- the outer, noisy regions of the image have been trimmed. The white 
contour is the 3-$\sigma$ level from the signal-to-noise 
image. Black contours show flux densities of 260, 440, 620, 800, 970, 1230 and
1500~mJy\,beam$^{-1}$.  Panels (a) and (b) are $310$ arcsec square while panels 
(c) and (d) are $285$ arcsec square. }
\label{fig:ngc660}
\end{figure*}

\begin{figure*}
\setlength{\unitlength}{1in}
\begin{picture}(7.0,5.5)
\includegraphics{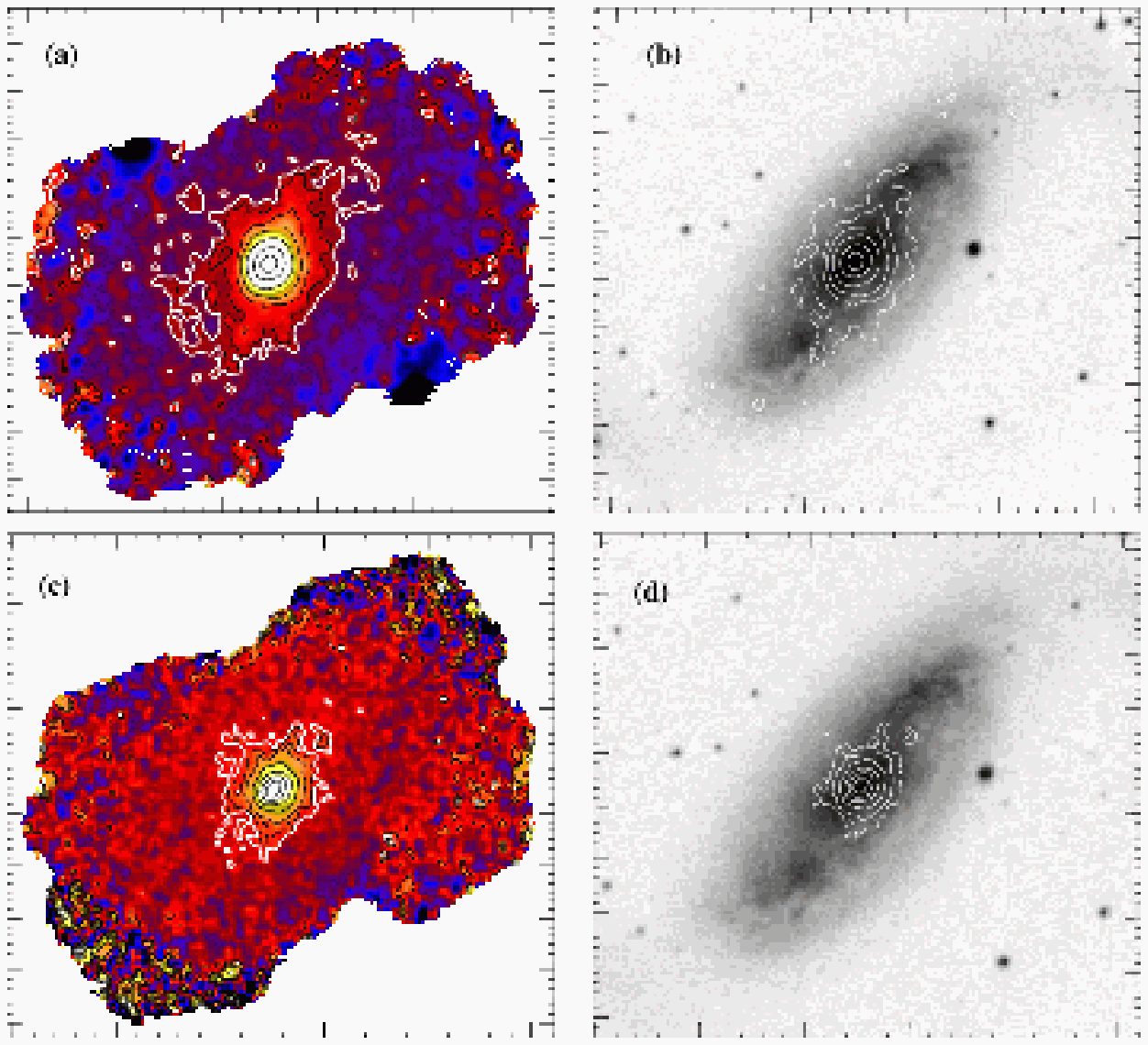}
\end{picture}
\caption[dum]{Images of NGC 1808 (a) 850-$\mu$m image as contoured colour scale. The
white contour is the 3-$\sigma$ level from the signal-to-noise image. Black
contours show flux densities of 15, 30, 50, 100, 200 and
300~mJy\,beam$^{-1}$. (b) DSS
image overlaid with the 850-$\mu$m contours. (c) 450-$\mu$m image as contoured
colour scale. The white 
contour is the 3-$\sigma$ level from the signal-to-noise 
image. Black contours show flux densities of 150, 250, 450, 700, 1000 and
1300~mJy\,beam$^{-1}$.  Panels (a) and (b) are $335\times310$ arcsec while panels 
(c) and (d) are $310\times285$ arcsec square.}
\label{fig:ngc1808}
\end{figure*}

\begin{figure*}
\setlength{\unitlength}{1in}
\begin{picture}(7.0,5.5)
\includegraphics{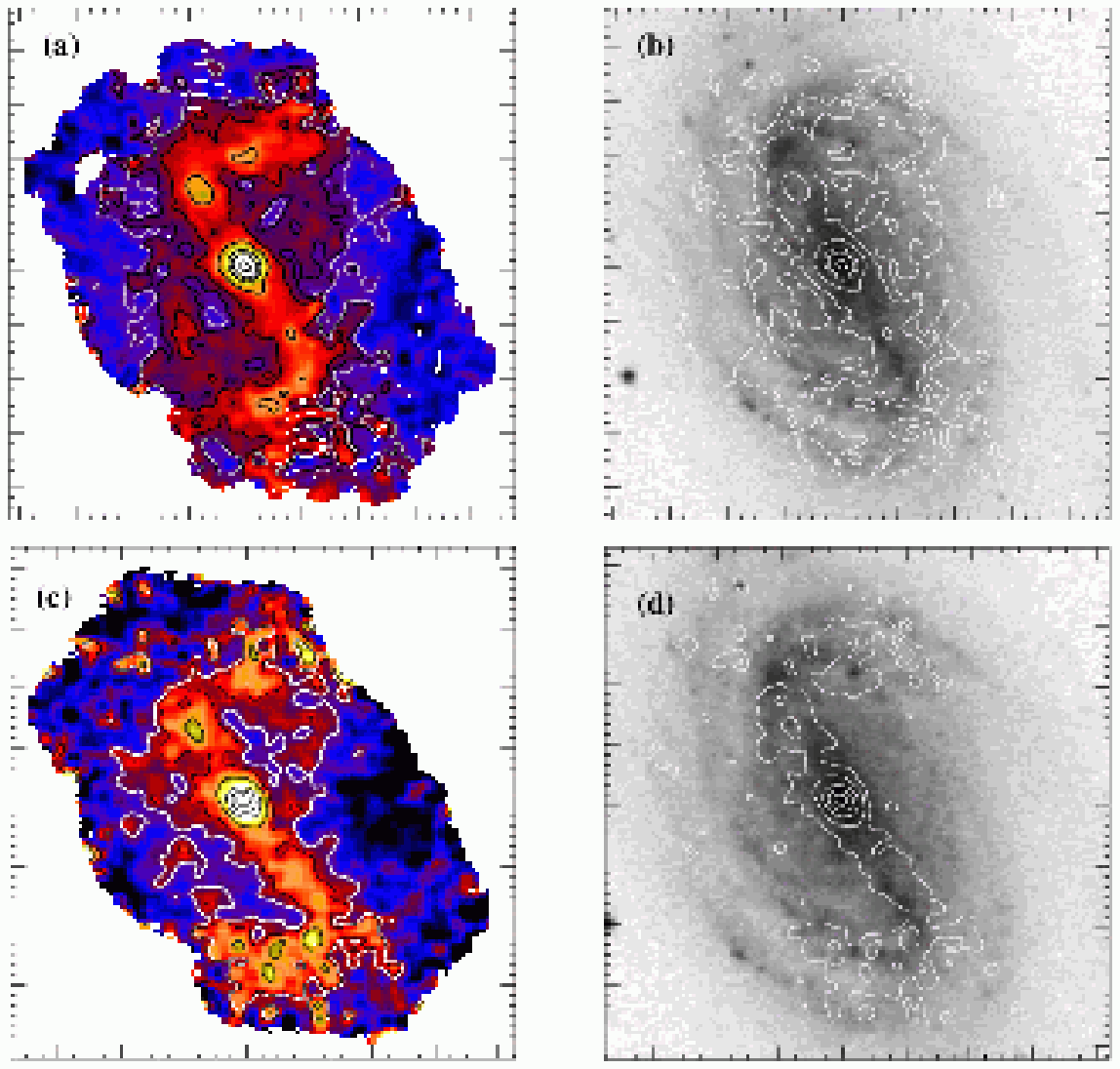}
\end{picture}
\caption[dum]{Images of NGC 2903 (a) 850-$\mu$m image as contoured colour scale. The
white contour is the 3-$\sigma$ level from the signal-to-noise image. Black
contours show flux densities of 15, 25, 50, 75, 100 and
125~mJy\,beam$^{-1}$. (b) DSS
image overlaid with the 850-$\mu$m contours. (c) 450-$\mu$m image as contoured
colour scale. The white contour is the 3-$\sigma$ level from the signal-to-noise 
image. Black contours show flux densities of 80, 150, 250, 350 and
450~mJy\,beam$^{-1}$.  Panels (a) and (b) are $245\times275$ arcsec while panels 
(c) and (d) are $220\times255$ arcsec.}
\label{fig:ngc2903}
\end{figure*}

\section{Results}

\subsection{Submillimetre images}




We present the submillimetre images in Figs~2--15. Two types of contour are
shown on each image. The white contour is taken from the S/N map and
overlaid on the signal map to allow the reader to assess the reality of any
faint, extended structure. Note that this procedure is necessary because the
noise often varies substantially across SCUBA maps (particularly at 450 $\mu$m) 
because of varying integration time from position to position and changing 
weather conditions. The black contours,
showing levels of equal flux density, are chosen to highlight
the extended source structure, any prominent spiral arms and/or bars and the
galactic cores. The maps are calibrated into Jy\,beam$^{-1}$ and are corrected for
contamination due to the error beam. These contours are overlaid on both the 
submillimetre maps and
Digitized Sky Survey (DSS) images.  The DSS images presented here are
from the blue-band second generation survey available from the Space Telescope
Science Institute and have a resolution of 1 arcsec.  

\begin{figure*}
\setlength{\unitlength}{1in}
\begin{picture}(7.0,3.8)
\includegraphics{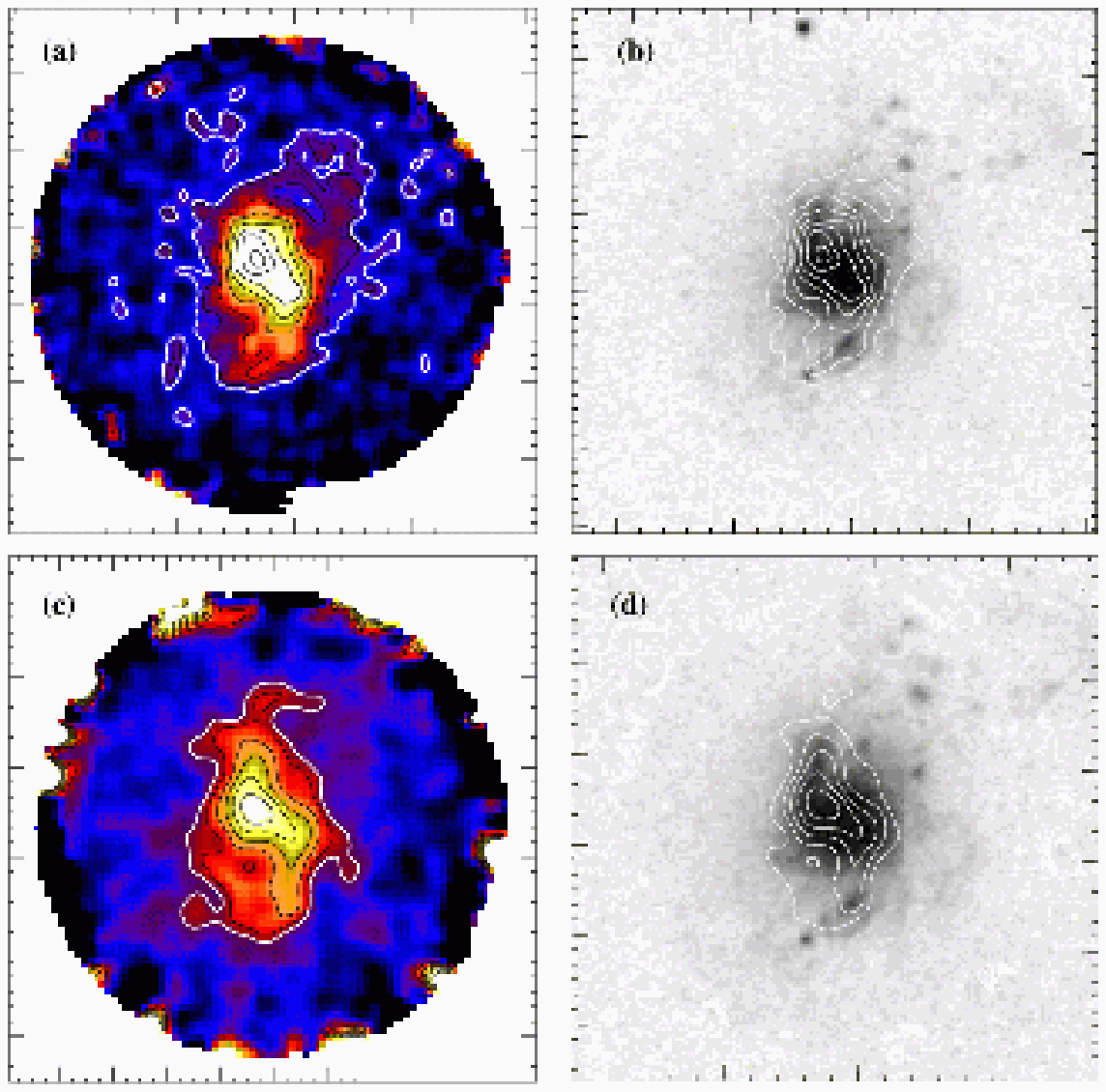}
\end{picture}
\caption[dum]{Images of NGC 3310 (a) 850-$\mu$m image as contoured colour scale. The
white contour is the 3-$\sigma$ level from the signal-to-noise image. Black
contours show flux densities of 6, 11, 16, 21, 26, 31, 36 and 41~mJy\,beam$^{-1}$. (b) DSS
image overlaid with the 850-$\mu$m contours. (c) 450-$\mu$m image as contoured
colour scale. The white contour is the 3-$\sigma$ level from the signal-to-noise 
image. Black contours show flux densities of 40, 60, 80, 100 and
120~mJy\,beam$^{-1}$. Panels (a) and (b) are 200 arcsec square while panels (c) and (d)
are 170 arcsec square.}
\label{fig:ngc3310}
\end{figure*}

\begin{figure*}
\setlength{\unitlength}{1in}
\begin{picture}(7.0,3.8)
\includegraphics{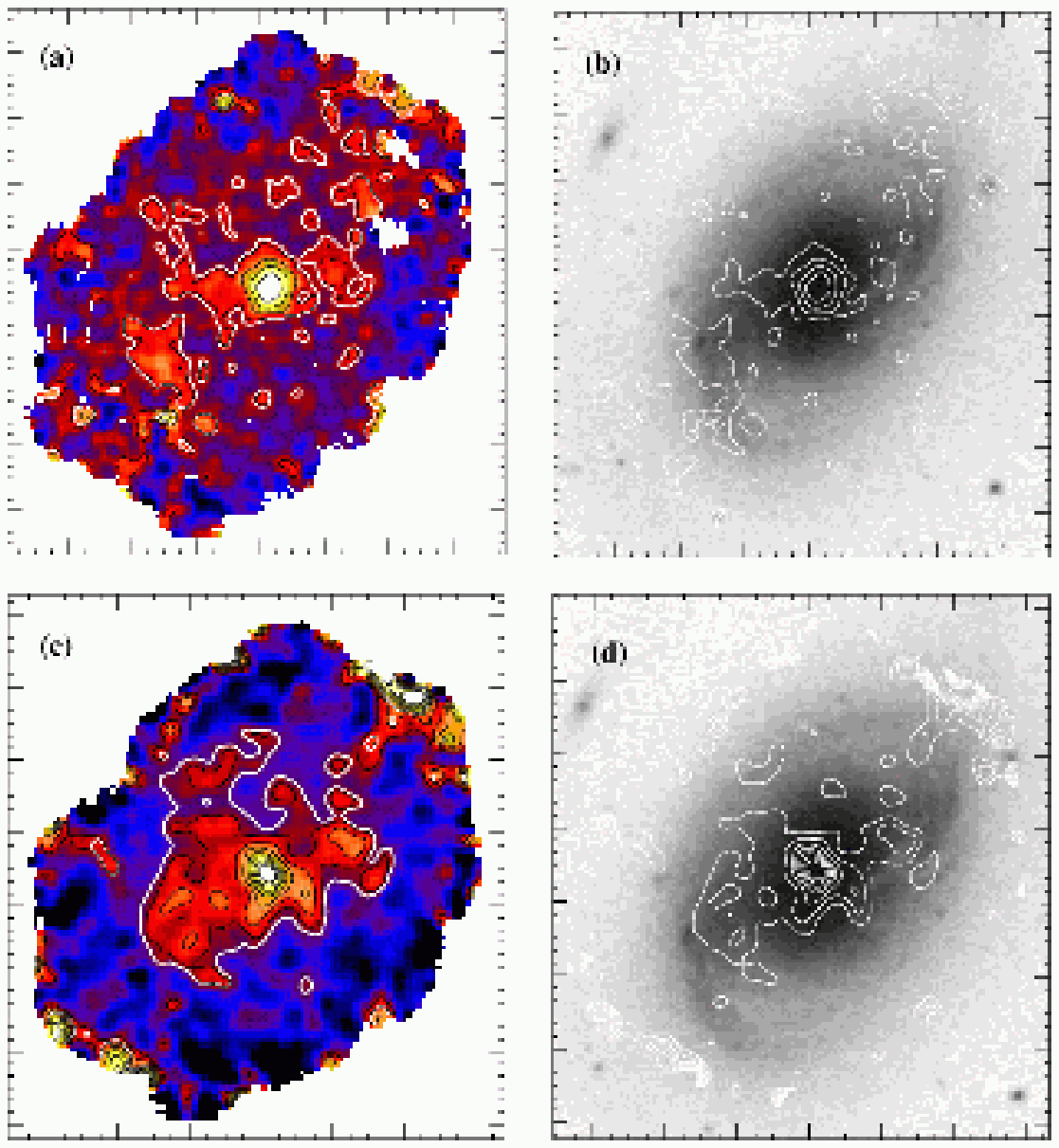}
\end{picture}
\caption[dum]{Images of NGC 3368 (a) 850-$\mu$m image as contoured colour scale. The
white contour is the 3-$\sigma$ level from the signal-to-noise image. Black
contours show flux densities of 15, 30, 50 and 70~mJy\,beam$^{-1}$. (b) DSS
image overlaid with the 850-$\mu$m contours. (c) 450-$\mu$m image as contoured
colour scale. The white contour is the 3-$\sigma$ level from the signal-to-noise 
image. Black contours show flux densities of 80, 110, 140, 170, 200, 230 and
260~mJy\,beam$^{-1}$. Panels (a) and (b) are $225\times245$ arcsec while panels 
(c) and (d) are $200\times220$ arcsec.}
\label{fig:ngc3368}
\end{figure*}

\begin{figure*}
\setlength{\unitlength}{1in}
\begin{picture}(7.0,7.5)
\includegraphics{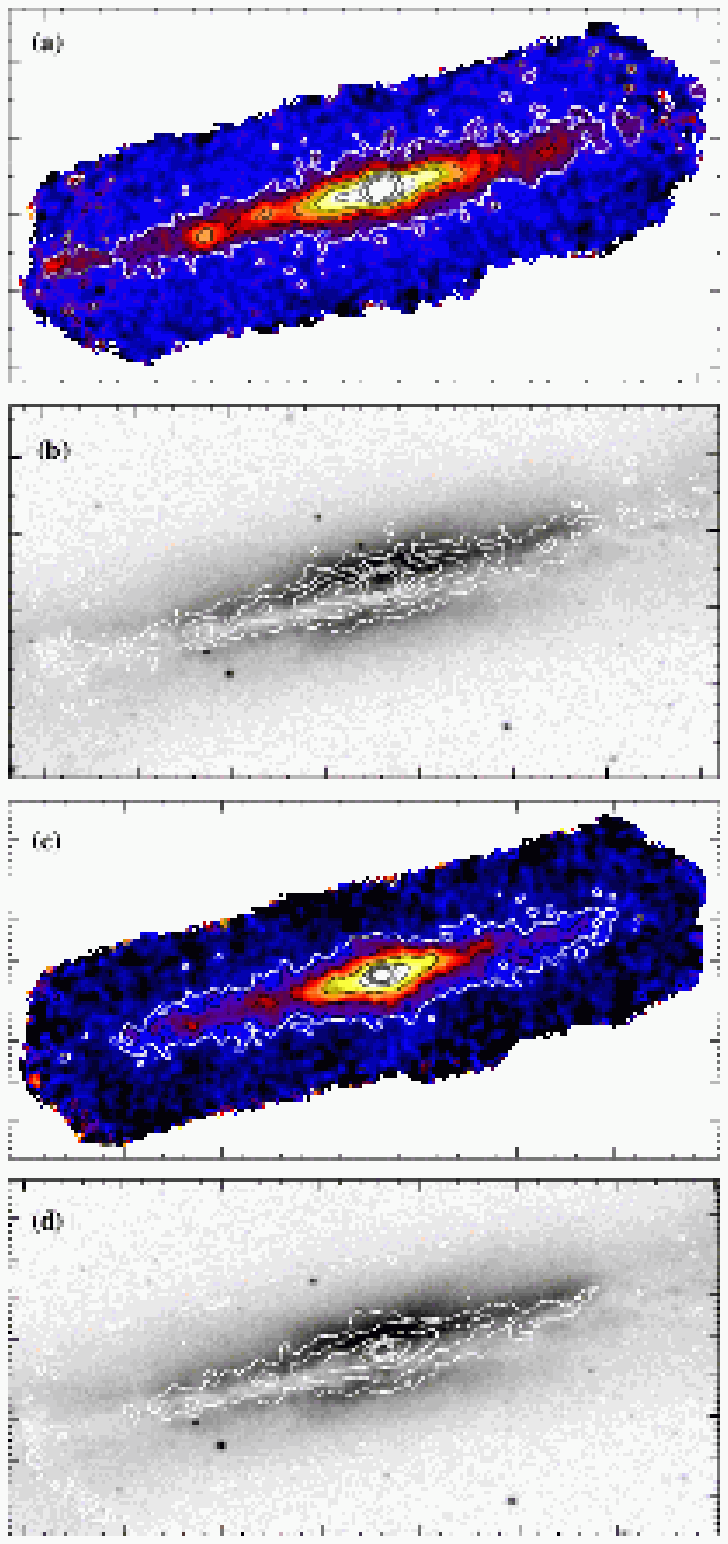}
\end{picture}
\caption[dum]{Images of NGC 3628 (a) 850-$\mu$m image as contoured colour scale. The
white contour is the 3-$\sigma$ level from the signal-to-noise image. Black
contours show flux densities of 15, 30, 50, 80, 110, 140 and 200~mJy\,beam$^{-1}$. (b) DSS
image overlaid with the 850-$\mu$m contours. (c) 450-$\mu$m image as contoured
colour scale. The white contour is the 3-$\sigma$ level from the signal-to-noise 
image. Black contours show flux densities of 100, 200, 400, 600, 800, 1000 and
1200~mJy\,beam$^{-1}$. Panels (a) and (b) are $550\times290$ arcsec while panels 
(c) and (d) are $525\times260$ arcsec.}
\label{fig:ngc3628}
\end{figure*}

\begin{figure*}
\setlength{\unitlength}{1in}
\begin{picture}(7.0,3.8)
\includegraphics{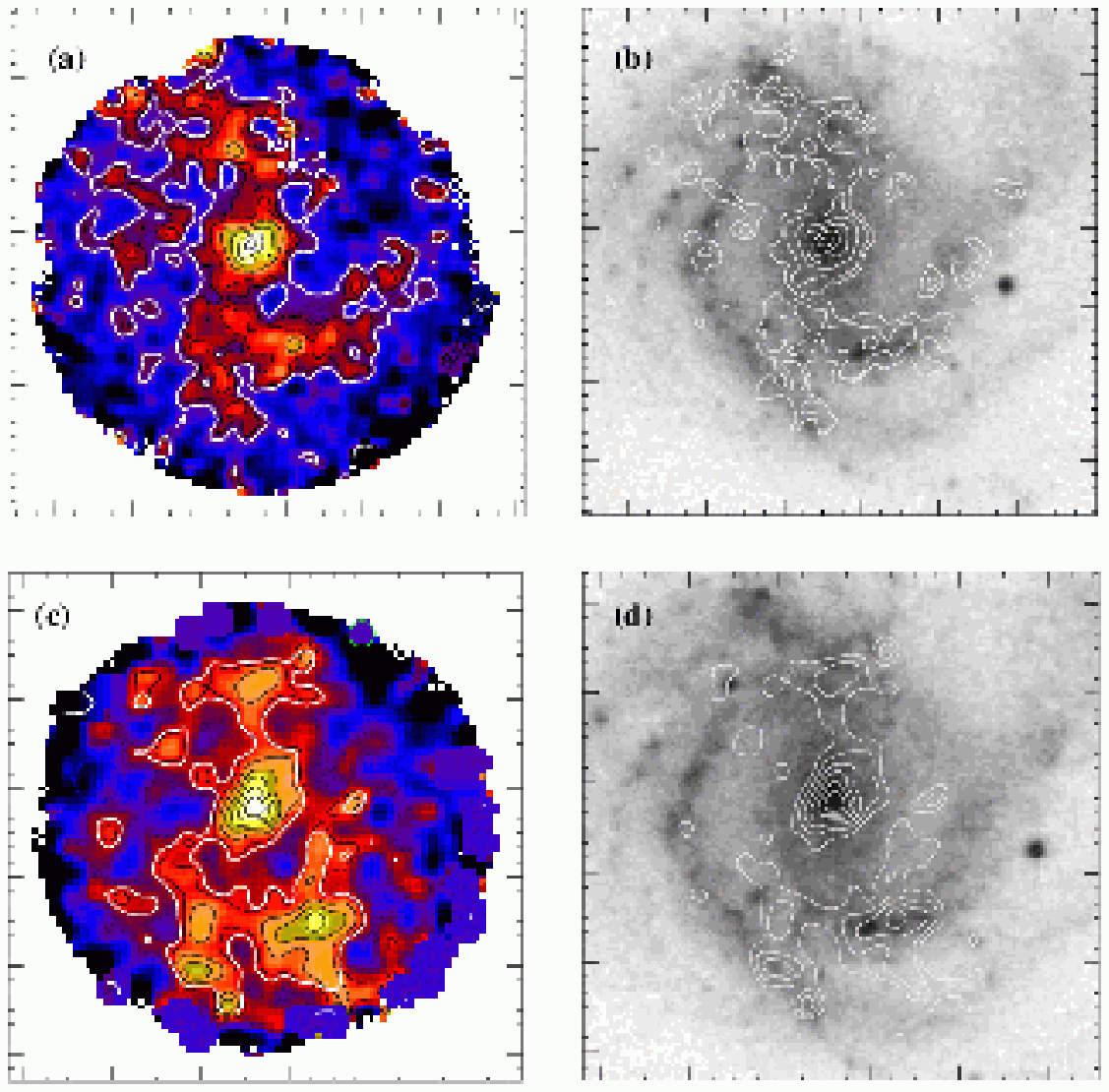}
\end{picture}
\caption[dum]{Images of NGC 4303 (a) 850-$\mu$m image as contoured colour scale. The
white contour is the 3-$\sigma$ level from the signal-to-noise image. Black
contours show flux densities of 18, 24, 36, 48, 60 and 72~mJy\,beam$^{-1}$. (b) DSS
image overlaid with the 850-$\mu$m contours. (c) 450-$\mu$m image as contoured
colour scale. The white contour is the 3-$\sigma$ level from the signal-to-noise 
image. Black contours show flux densities of 70, 90, 110, 130, 150 and
170~mJy\,beam$^{-1}$. Panels (a) and (b) are $195$ arcsec square while panels 
(c) and (d) are $170$ arcsec square.}
\label{fig:ngc4303}
\end{figure*}

\begin{figure*}
\setlength{\unitlength}{1in}
\begin{picture}(7.0,2.0)
\includegraphics{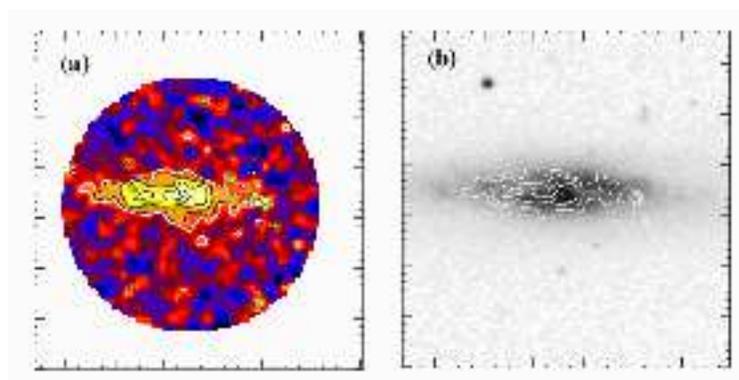}
\end{picture}
\caption[dum]{Images of NGC 4388 (a) 850-$\mu$m image as contoured colour scale. The
white contour is the 3-$\sigma$ level from the signal-to-noise image. Black
contours show flux densities of 20, 30, 40 and 50~mJy\,beam$^{-1}$. (b) DSS
image overlaid with the 850-$\mu$m contours. The panels are $195$ arcsec square.}
\label{fig:ngc4388}
\end{figure*}

\begin{figure*}
\setlength{\unitlength}{1in}
\begin{picture}(7.0,3.8)
\includegraphics{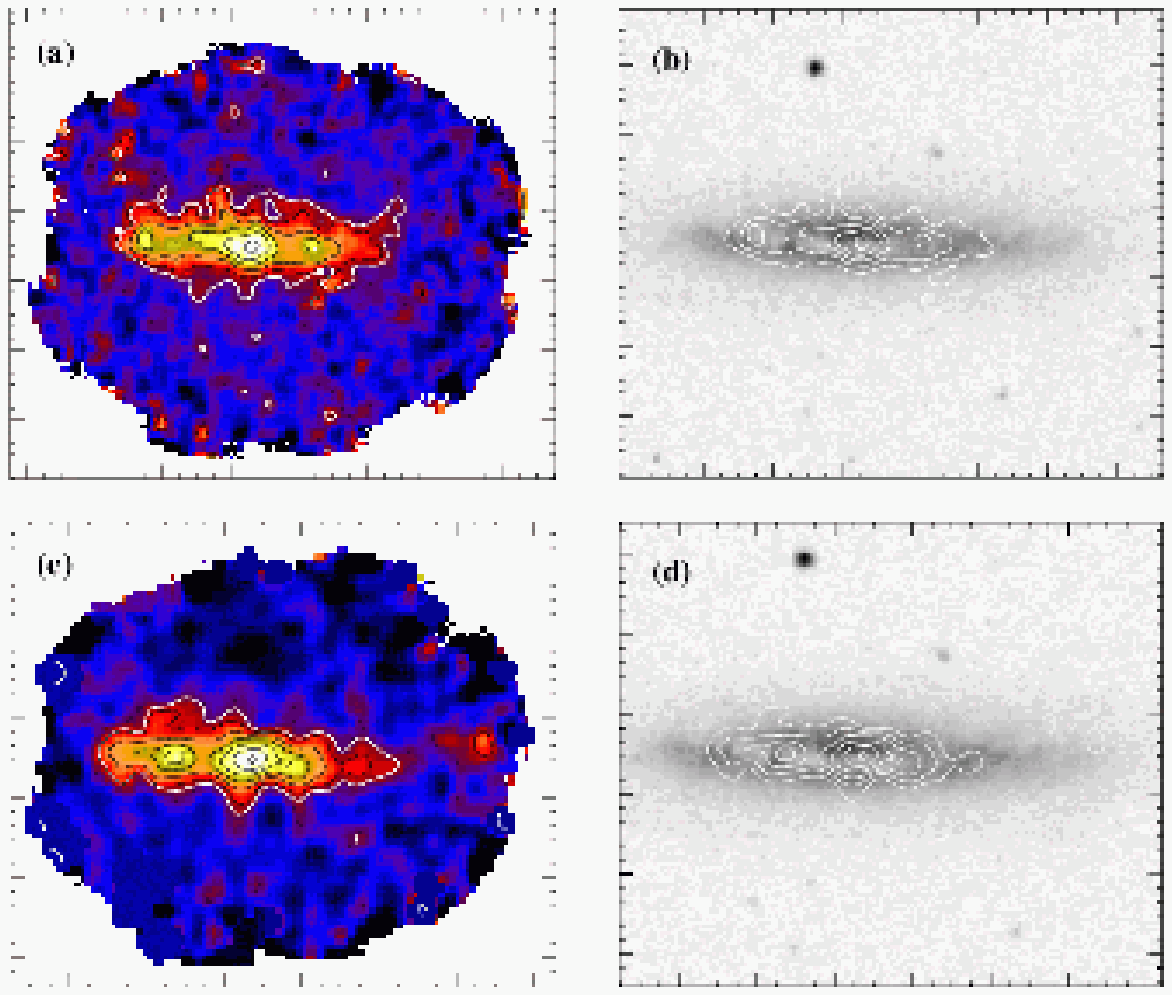}
\end{picture}
\caption[dum]{Images of NGC 4402 (a) 850-$\mu$m image as contoured  colour scale. The
white contour is the 3-$\sigma$ level from the signal-to-noise image. Black
contours show flux densities of 15, 25, 35, 45 and 55~mJy\,beam$^{-1}$. (b) DSS
image overlaid with the 850-$\mu$m contours. (c) 450-$\mu$m image as contoured
colour scale. The white contour is the 3-$\sigma$ level from the signal-to-noise 
image. Black contours show flux densities of 70, 100, 130, 160, 190 and
220~mJy\,beam$^{-1}$. Panels (a) and (b) are $230\times200$ arcsec while panels 
(c) and (d) are $200\times170$ arcsec.}
\label{fig:ngc4402}
\end{figure*}

\begin{figure*}
\setlength{\unitlength}{1in}
\begin{picture}(7.0,3.8)
\includegraphics{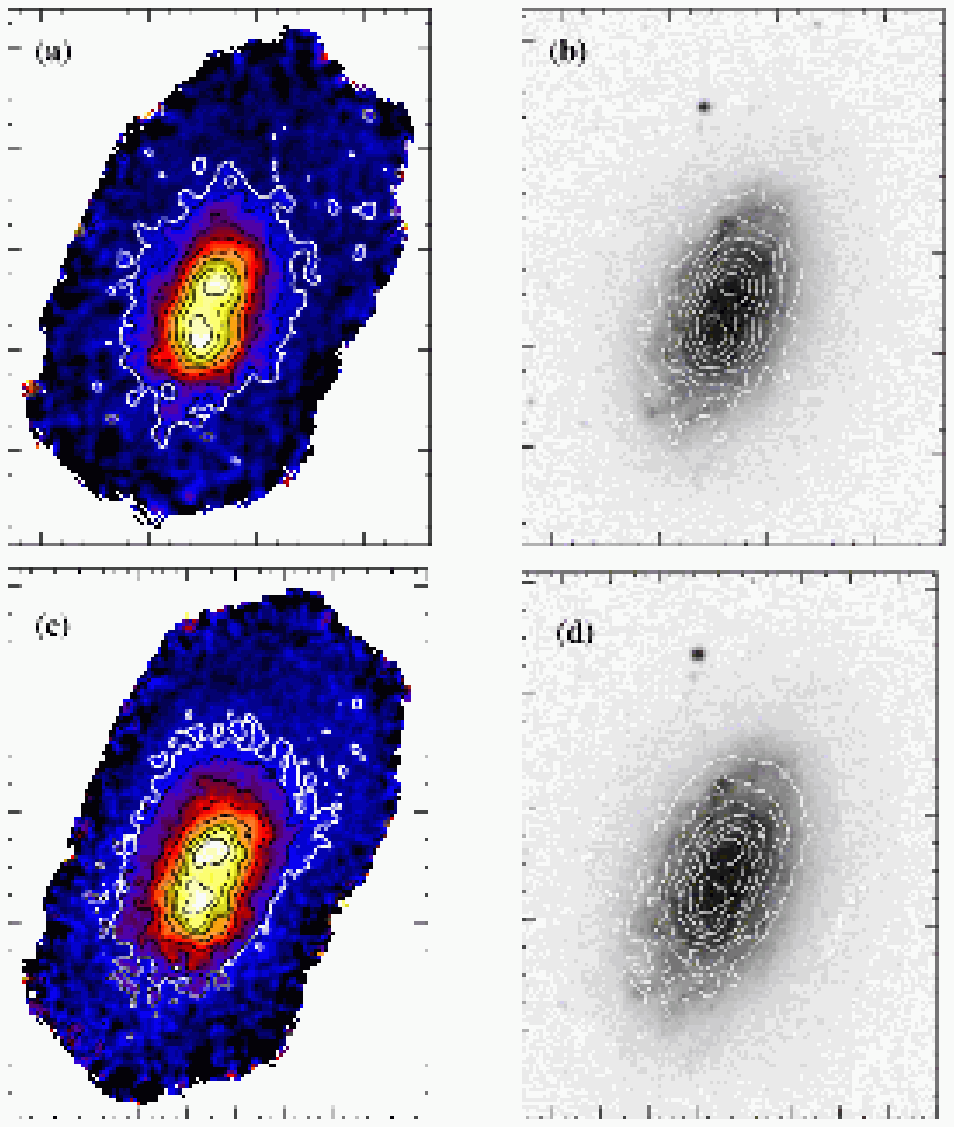}
\end{picture}
\caption[dum]{Images of NGC 4414 (a) 850-$\mu$m image as contoured colour scale. The
white contour is the 3-$\sigma$ level from the signal-to-noise image. Black
contours show flux densities of 10, 20, 30, 40, 50, 60 and 70~mJy\,beam$^{-1}$. (b) DSS
image overlaid with the 850-$\mu$m contours. (c) 450-$\mu$m image as contoured
colour scale. The white contour is the 3-$\sigma$ level from the signal-to-noise 
image. Black contours show flux densities of 40, 80, 120, 160, 200, 240 and
280~mJy\,beam$^{-1}$. Panels (a) and (b) are $250\times315$ arcsec while panels 
(c) and (d) are $220\times310$ arcsec.}
\label{fig:ngc4414}
\end{figure*}

\begin{figure*}
\setlength{\unitlength}{1in}
\begin{picture}(7.0,4.2)
\includegraphics{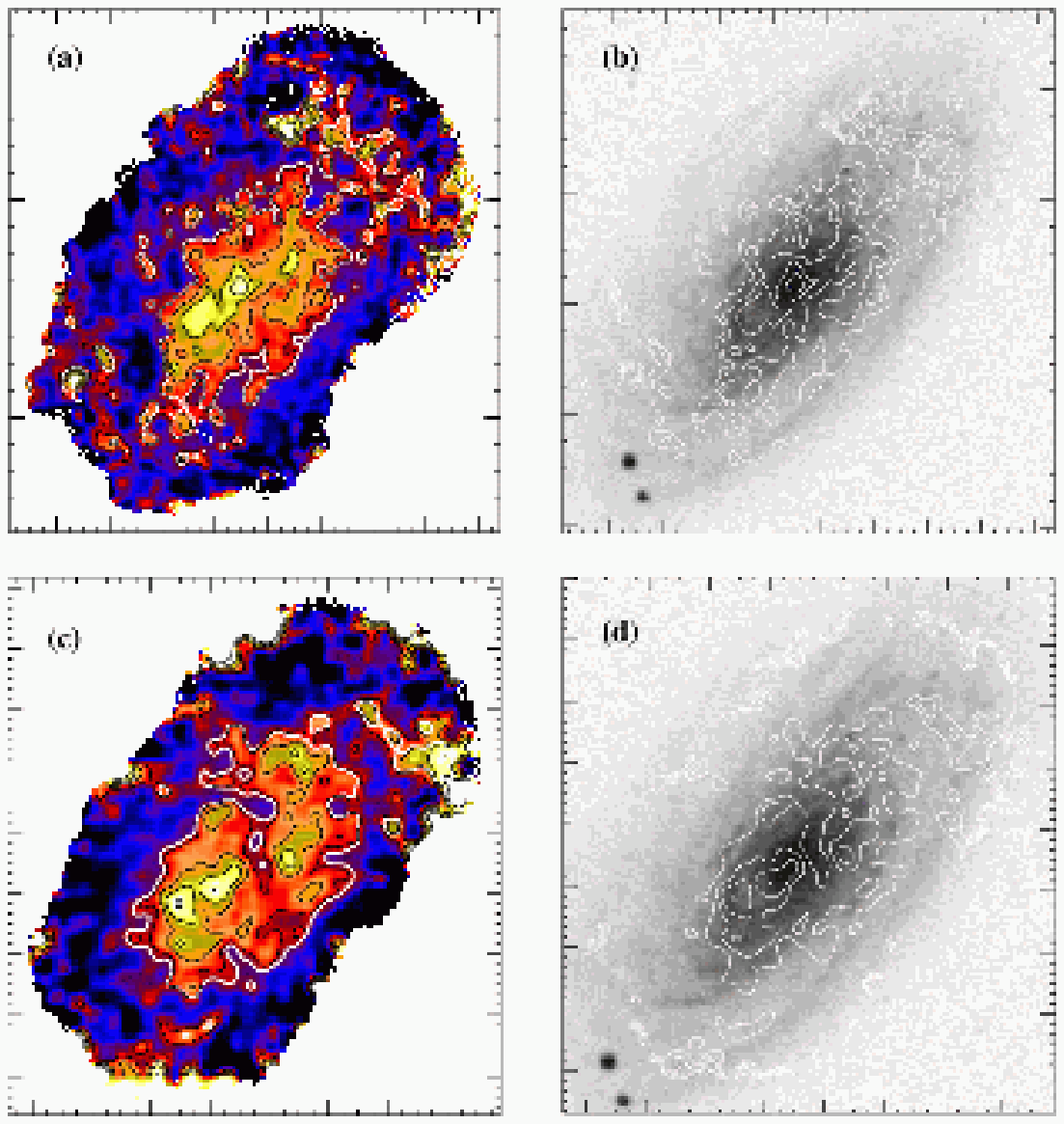}
\end{picture}
\caption[dum]{Images of NGC 4501 (a) 850-$\mu$m image as contoured colour scale. The
white contour is the 3-$\sigma$ level from the signal-to-noise image. Black
contours show flux densities of 15, 25, 35 and 45~mJy\,beam$^{-1}$. (b) DSS
image overlaid with the 850-$\mu$m contours. (c) 450-$\mu$m image as contoured
colour scale. The white contour is the 3-$\sigma$ level from the signal-to-noise 
image. Black contours show flux densities of 80, 120, 160 and
200~mJy\,beam$^{-1}$. Panels (a) and (b) are $265\times265$ arcsec while panels 
(c) and (d) are $240\times260$ arcsec.}
\label{fig:ngc4501}
\end{figure*}

\begin{figure*}
\setlength{\unitlength}{1in}
\begin{picture}(7.0,7.5)
\includegraphics{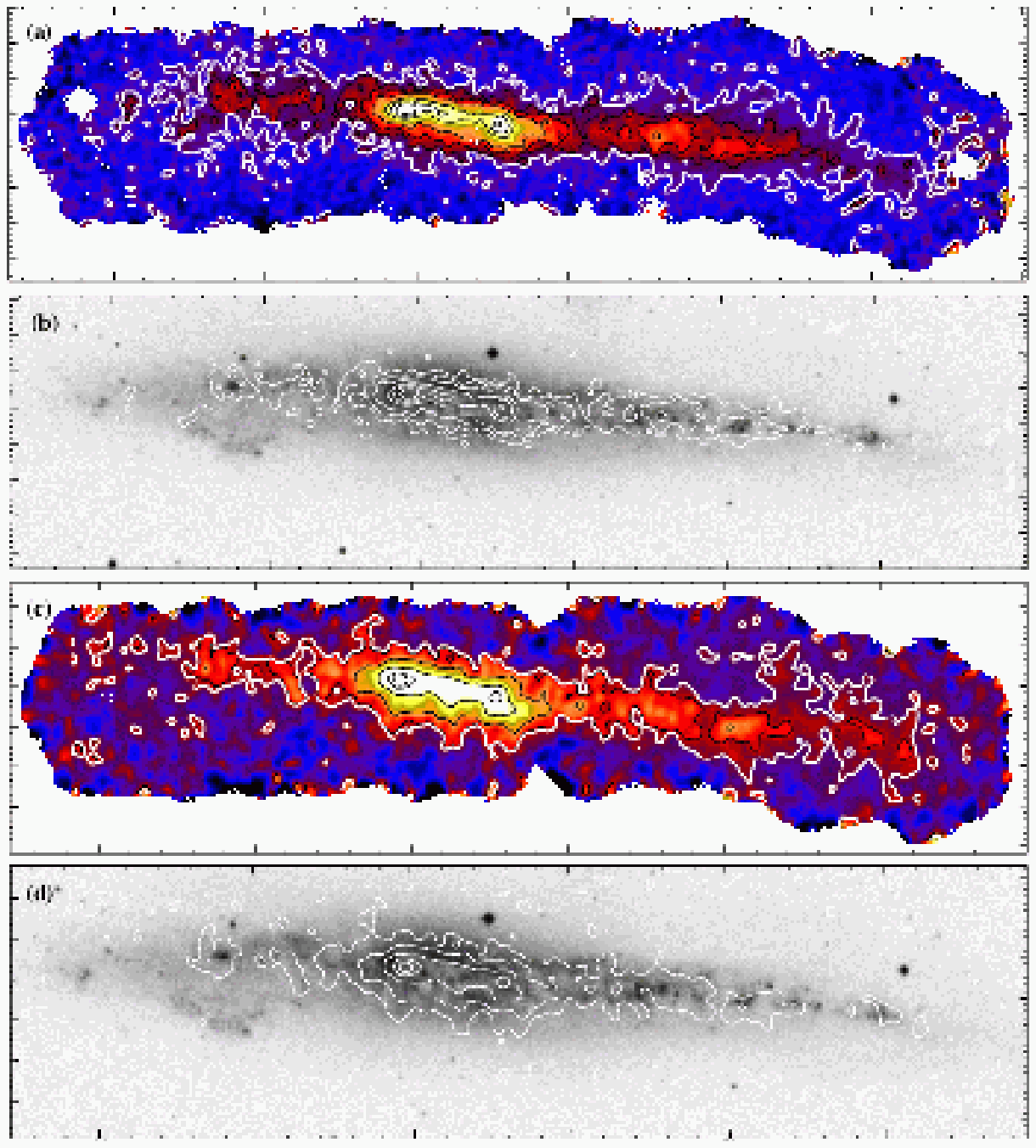}
\end{picture}
\caption[dum]{Images of NGC 4631 (a) 850-$\mu$m image as contoured  colour scale. The
white contour is the 3-$\sigma$ level from the signal-to-noise image. Black
contours show flux densities of 20, 30, 45, 80, 100 and 120~mJy\,beam$^{-1}$. (b) DSS
image overlaid with the 850-$\mu$m contours. (c) 450-$\mu$m image as contoured
colour scale. The white contour is the 3-$\sigma$ level from the signal-to-noise 
image. Black contours show flux densities of 50, 100, 200, 300 and
350~mJy\,beam$^{-1}$.  Panels (a) and (b) are $845\times225$ arcsec while panels 
(c) and (d) are $820\times200$ arcsec.}
\label{fig:ngc4631}
\end{figure*}

\begin{figure*}
\setlength{\unitlength}{1in}
\begin{picture}(7.0,4.2)
\includegraphics{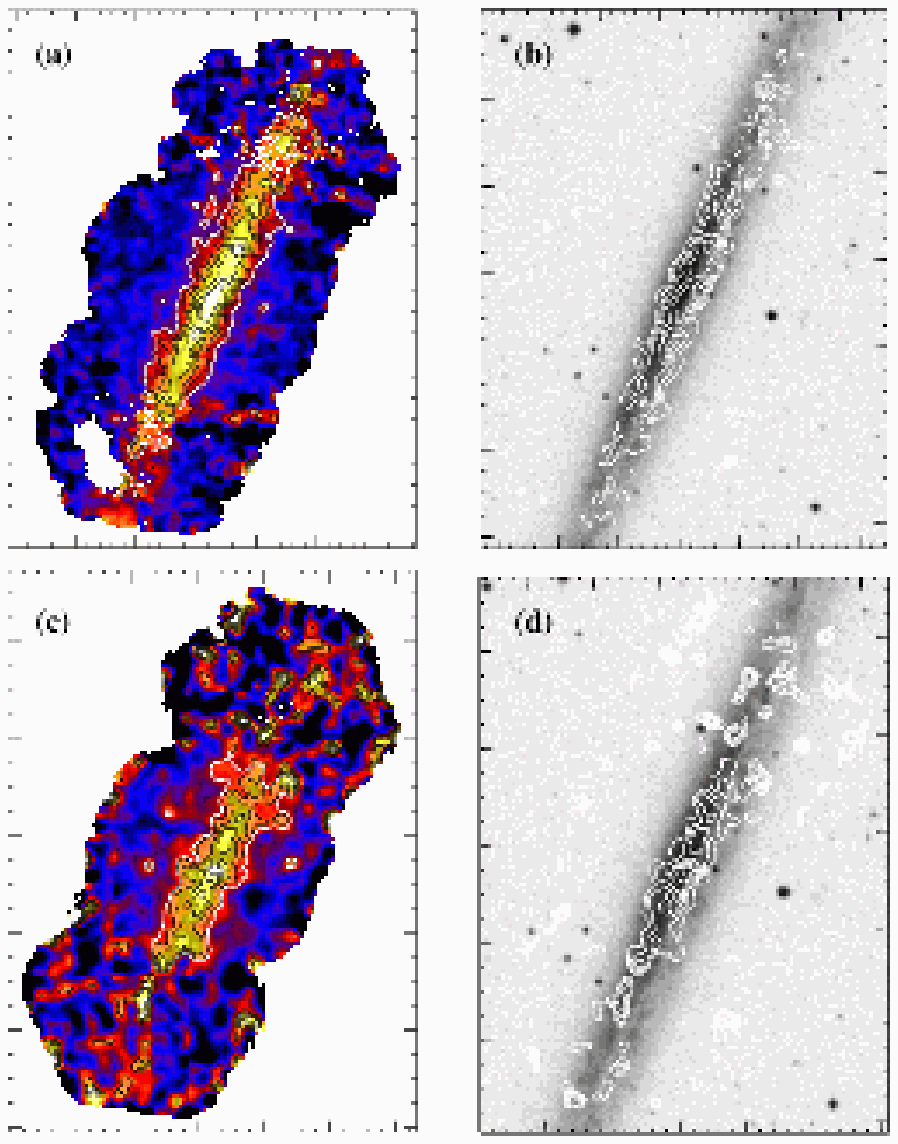}
\end{picture}
\caption[dum]{Images of NGC 5907 (a) 850-$\mu$m image as contoured colour scale. The
white contour is the 3-$\sigma$ level from the signal-to-noise image. Black
contours show flux densities of 15, 20, 25, 30 and 35~mJy\,beam$^{-1}$. (b) DSS
image overlaid with the 850-$\mu$m contours. (c) 450-$\mu$m image as contoured
colour scale. The white contour is the 3-$\sigma$ level from the signal-to-noise 
image. Black contours show flux densities of 150, 200, 250, 300 and
350~mJy\,beam$^{-1}$. Panels (a) and (b) are $280\times370$ arcsec while panels 
(c) and (d) are $250\times345$ arcsec.}
\label{fig:ngc5907}
\end{figure*}

\subsubsection{Comments on individual objects}

{\bf NGC~157}\\ A prominent thick spiral structure is seen in both the
submillimetre maps for this partially inclined galaxy (i = 49$^{\circ}$).  A
high contrast between the arm and inter-arm regions is seen. The peaks of the
submillimetre maps are centred on (or very near to) dark lanes within the grand
spiral arms of the optical image.

{\noindent\bf NGC~660}\\ This inclined galaxy (i = 66$^{\circ}$) is core
dominated in both the submillimetre optical maps.  The submillimetre peak is
offset from the optical peak by $\sim$ 7 arcsec to the north-west,
corresponding to a projected linear offset of $\sim$ 670~pc. However, the
intrinsic centre of the distribution of stars (and therefore starlight) in the
galaxy is likely to be given by the submillimetre peak.  The apparent optical
peak may simply be the brightest part of the galactic core that is not obscured
by dust.

{\noindent\bf NGC~1808}\\ The submillimetre maps of this partially inclined
galaxy ($i = 60^{\circ}$) are very core dominated.  However, the contours are
seen to be elongated along the major-axis on the higher S/N 850-$\mu$m map.
There is no evidence of the unusual small bar seen in the optical image,
approximately $20^{\circ}$ offset from the major axis.  Even though the optical
image is dominated by the core, the starlight from this region may be heavily
attenuated by the highly concentrated dust.

{\noindent\bf NGC~2903}\\ The bar of this spectacular spiral is clearly seen at
both 450 and 850~$\mu$m.  The $\sim$125 arcsec submillimetre bar corresponds to
5.5~kpc at the assumed distance of NGC~2903.  Knots of bright submillimetre
emission are seen along the length of the bar as well as along the spiral arms
attached to the bar ends.  There is a very high submillimetre contrast between
the bar/arm regions and the inter-arm regions and a bright unresolved galactic
core is seen in the submillimetre. There is some danger that we may have
chopped some weak extended emission out of this image.

{\noindent\bf NGC~3310}\\ Submillimetre emission is seen out to $\sim$2/3 of
the optical radius in most directions on the 850-$\mu$m map.  The peak of the
submillimetre emission is offset by $\sim$12 arcsec to the north-east from the
centre of the optical core.  This corresponds to a linear distance of 1.2~kpc
at the assumed distance of NGC~3310.

{\noindent\bf NGC~3368}\\ This mid-inclination galaxy ($i = 45^{\circ}$) has a
least one short bar, which is very well picked out by the submillimetre images.
The unresolved core is relatively bright at both submillimetre wavelengths,
however the core to bar luminosity ratio appears to be much higher on the
850-$\mu$m map, suggesting either the bar contains warm dust or (less likely)
the core contains cool dust.

{\noindent\bf NGC~3628}\\ The optical image of this edge-on galaxy is heavily
obscured by a prominent dust lane.  The submillimetre emission follows the dust
lane extremely well, however it is seen slightly to the north.  This effect is
caused by the slight inclination of the galaxy on the sky; as the observed
submillimetre emission samples both the southern near-side associated with the
dust lane, and the unseen northern far side. 

{\noindent\bf NGC~4303}\\ The grand-design spiral arms of this face-on Virgo
spiral ($i = 25^{\circ}$) are mapped at the ends of a north-south running bar
in the submillimetre images, especially at 850~$\mu$m.  As with NGC~2903,
bright knots of emission are seen along the bar and arms.  There is a strong
contrast between the arm and inter-arm regions but the core is still the
brightest submillimetre region of NGC~4303. There is some danger that we may
have chopped some weak extended emission out of this image.

{\noindent\bf NGC~4388}\\ Unfortunately this edge-on Virgo spiral ($i =
79^{\circ}$) was observed under rather poor weather conditions and was not
detected at 450~$\mu$m.  The S/N of the 850-$\mu$m is much better; three
submillimetre peaks are seen, the central one coincides with the brightest
optical region, the eastern peak is offset by $\sim 28$ arcsec and the weaker
western peak by $\sim$37 arcsec.  These angular distances correspond to 2.7~kpc
and 3.5~kpc respectively at the assumed distance to the Virgo cluster.  They
may be explained as the increased column density seen through a spiral arm on
each side of the galactic core.  The submillimetre core itself is only slightly
brighter than the eastern peak.

{\noindent\bf NGC~4402}\\ This is a second edge-on Virgo spiral ($i =
75^{\circ}$), however its optical appearance differs from NGC~4388 as it has a
prominent dust lane curving through it.  Once again it has three submillimetre
peaks, with the central one the brightest, closely followed by the eastern, and
then finally the western peak which is barely recognizable as above the disc
emission.

{\noindent\bf NGC~4414}\\ This mid-inclination spiral ($i = 54^\circ$) is
thought to be in front of the Virgo cluster.  It has two central peaks which
are best explained as the enhanced line-of-sight column densities seen through
a ring orbiting the galactic core.  The $\sim$30 arcsec separation of the peaks
corresponds to a linear distance of 2.1~kpc at the assumed distance to
NGC~4414.  The idea of a dusty ring surrounding the core is supplemented by the
molecular gas hole detected in the same region by Sakamoto et~al. (1999).

{\noindent\bf NGC~4501}\\ Unfortunately both SCUBA bolometer arrays were so
noisy throughout the observations of this mid-inclination Virgo spiral ($i =
58^\circ$) that 25 per cent of the 450-$\mu$m bolometers and 40 per cent of the
850-$\mu$m bolometers had to be blanked.  Nonetheless, submillimetre emission
is detected from the galaxy and its peak coincides with the optical centre.

{\noindent\bf NGC~4631}\\ The submillimetre emission from this large galaxy ($i
= 85^\circ$)has been mapped over almost 14~arcmin which corresponds to a linear
distance of 52~kpc at the assumed distance to NGC~4631. This is 92 per cent of
the optical radius out to the $25^{\rm th}$ mag isophote.  Several features are
seen in the submillimetre maps including, three central peaks, a warp in the
western side of the disc, knots of bright emission along the disc and
filamentary structure emanating like smoke from chimneys away from the plane of
the disc.

The three central peaks may be explained from a dusty torus surrounding the
centre of the galaxy, as seen in the edge-on galaxies NGC~4388 and NGC~4402.
The $\sim 80$ arcsec separation of the two outer peaks correspond to a diameter
of 4.9~kpc at the assumed distance to NGC~4631.  The eastern peak is brighter
than the western one, however in this case both are brighter than the central
peak.  The 850-$\mu$m emission, and arguably the 450-$\mu$m emission, is seen
to peak twice more $\sim 90$ arcsec either side of the central peak,
corresponding to a linear distance of 5.5~kpc.  These peaks may be due to the
increased line of sight column densities through two main spiral arms.

{\noindent\bf NGC~5907}\\ The galaxy is almost perfectly edge-on
($i=89^\circ$).  The majority of the submillimetre emission is extremely well
confined to a thin disc and is well mapped out to just over half the optical
radius.  It is seen to be dominated by the core which is unresolved by SCUBA in
the direction perpendicular to the galactic plane.  A few bright knots of
submillimetre emission are seen along the disc.  The two peaks $\sim 75$ arcsec
either side of the centre correspond to a linear distance of 5.7~kpc at the
assumed distance to the galaxy.  A third, more distant northern peak seen on
the 850-$\mu$m map is 120 arcsec from the centre, or 9.1~kpc.  All three may be
due to the increased dust column density along the line of sight through spiral
arm regions within the galactic disc.

\subsection{Flux densities}

\begin{table*}
\begin{minipage}{125mm}

\caption{Submillimetre (this work) and far-infrared ({\em IRAS\/}) flux densities.}
\label{table:fluxes}
\begin{tabular}{lcccccc}\hline
(1) &(2)& (3)& (4)& (5)& (6)& (7)\\
& \multicolumn{2}{c}{Flux density to $1\sigma$ contour} &
\multicolumn{2}{c}{Flux density in map}&
\multicolumn{2}{c}{{\em IRAS\/} far-infrared flux densities}\\ 
Source & $S_{850}$ & $S_{450}$ & $S_{850}$ & $S_{450}$ &$S_{100}$ & $S_{60}$\\ 
& (mJy) & (mJy) & (mJy) & (mJy) & (Jy) & (Jy) \\\hline
NGC 157 & $910\pm13$ & $4734\pm81$ & \ldots & \ldots & $43.1\pm4.3^a$ &
$17.6\pm1.8^a$ \\
NGC 660 & $1360\pm12$ & $6925\pm186$ & $1393\pm20$ & $8832\pm662$&
$104.9\pm10.5^b$ & $67.3\pm6.7^b$\\
NGC 1808& $1301\pm22$ & $7138\pm135$ & \ldots & $8133\pm271$& $137.2\pm5.5^c$ & $87.8\pm3.5^c$\\
NGC 2903& $1796\pm26$ & $7936\pm232$ & \ldots & \ldots& $147.4\pm22.1^b$ & $52.4\pm7.9^b$ \\
NGC 3310& $358\pm8$ & $1293\pm52$ & \ldots & $1406\pm71$& $48.0\pm4.8^a$ &
$34.1\pm3.4^a$ \\
NGC 3368& $513\pm23$ & $3439\pm120$ & \ldots & \ldots& $30.4\pm3.0^a$ & $10.7\pm1.1^a$\\
NGC 3628& $2822\pm184$ & $16934\pm196$ & \ldots & \ldots& $106.0\pm10.6^a$ & $51.6\pm5.2^a$\\ 
NGC 4303& $705\pm20$ & $4602\pm172$ & \ldots & \ldots& $79.7\pm8.0^a$ & $37.5\pm3.8^a$\\
NGC 4388& $234\pm13$ & \ldots & $294\pm21$ & \ldots& $17.4\pm1.7^a$ & $10.1\pm1.0^a$\\
NGC 4402& $357\pm10$ & $2309\pm94$ & $445\pm21$ & $2772\pm169$& $17.5\pm1.8^a$ & $5.4\pm0.5^a$\\
NGC 4414& $1160\pm12$ & $7543\pm60$ & \ldots & \ldots& $69.1\pm6.9^a$ & $30.1\pm3.0^a$\\
NGC 4501& $962\pm27$ & $6386\pm186$ & \ldots & \ldots& $63.7\pm6.4^a$ & $17.6\pm1.8^a$\\
NGC 4631& $5253\pm46$ & $25056\pm1731$ & \ldots & \ldots & $170.4\pm34.1^d$ & $90.7\pm18.2^d$ \\
NGC 5907& $1962\pm33$ & $13188\pm908$ & \ldots & \ldots& $45.8\pm6.9^b$ & $8.8\pm1.3^b$\\
\hline
\end{tabular}

(1) Source name. (2) and (3) 850- and 450-$\mu$m flux density calculated by
summing the emission to the 1-$\sigma$ contour (see text for more details).
(4) and (5) Total 850- and 450-$\mu$m flux density in the map if different from
that within the 1-$\sigma$ contour.  The quoted uncertainties represent the
noise in the given aperture placed on the noise maps. The total error is
dominated by the calibration uncertainty which we estimate to be $\sim20$ per
cent at 850~$\mu$m and $\sim25$ per cent at 450~$\mu$m (see text). (6) and (7)
100 and $60~\mu$m flux density from {\em IRAS\/}. These values are used in the
model fits. Literature sources are: $a$ -- Soifer et al. (1989); $b$ -- Rice et
al. (1988); $c$ -- Moshir et al. (1990); $d$ -- Young et al. (1989).

\end{minipage}
\end{table*}

We present total submillimetre flux densities in Table~\ref{table:fluxes}. For
each galaxy we made signal-to-noise maps and determined the 1-$\sigma$ contour
level. This contour could be approximated by an elliptical aperture.  Flux
densities were calculated by summing the emission in these apertures placed on
the signal maps, and uncertainties by adding in quadrature the data in the
corresponding aperture placed on the noise map. Calibration uncertainties were
calculated from the variation in the sky opacity and variation of FCF across
the night when the data was taken.

Some of the galaxies have low surface brightness emission extending across the
whole map. We estimated this additional emission for each galaxy by repeating
the above procedure and simply summing the total emission on the map. Using
this method, we find that five galaxies have higher flux densities at one or
both wavelengths. In addition, if the optical radius of the galaxy exceed our 2
arminute chop throw there is a danger of missing very extended flux. However
this is a danger only for NGC~2903 and NGC~4303, both of which appear to have
their emission strongly dominated by their prominent bars.

Global 60- and 100-$\mu$m {\em IRAS\/} flux densities were taken from the
literature. The primary source was data presented for the {\em IRAS\/} Bright
Galaxy Sample by Soifer et al. (1989), although 4 objects have flux densities
taken from other publications. These far-infrared flux densities, all
used in subsequent modelling, are listed in Table~\ref{table:fluxes} along
with references to their literature sources. We supplemented these values with
published data at 1300~$\mu$m for NGC~660 (Chini et al. 1986); at $170~\mu$m
for NGC~4388 and NGC~4402 (Tuffs et al. 2002); at 180$~\mu$m (Bendo et
al. 2003), 870 and 1230~$\mu$m (Dumke, Krause \& Wielebinski 2004) for
NGC~4631; and at 1200~$\mu$m (Dumke et al. 1997) for NGC~5907. We also include
our previously published data for NGC~3079 (Stevens \& Gear 2000) together with
the revised ISOPHOT data presented by Klaas \& Walker (2002), and include this
galaxy in subsequent analysis. For NGC~5907, Alton et al. (2004) report a total
flux density of 1.6 Jy ($\pm15$ per cent) at 850~$\mu$m which is consistent
with our value within the quoted uncertainties.

\subsection{Model fitting the spectral energy distributions}

\label{section:mod}

In this section we fit simple models to the spectral energy distributions
(SEDs) to estimate the temperatures and masses of the dust components in our
spiral galaxies.  Although, the dust is likely to be radiating at a range of
temperatures, simple two component models are often chosen to describe the
far-infrared--millimetre SEDs of galaxies. The choice is partly motivated by
physical considerations and partly demanded by lack of the data required to
constrain more complex models. Physically one might identify the warm component
with dust heated predominantly by OB stars and the cold component with dust in
`cirrus' clouds heated by the interstellar radiation field (ISRF).

We choose to fit the data with two optically thin greybody functions, giving an
equation of the form,
\begin{equation}
S(\lambda) = N_{\rm w} B(\lambda ,T_{\rm w}) Q_{\rm em}(\lambda) + N_{\rm c}
B(\lambda ,T_{\rm c}) Q_{\rm em}(\lambda) 
\end{equation}
where $S$ is the flux density at wavelength $\lambda$, $N_{\rm w}$ and $N_{\rm c}$ are the 
normalizations, $B(\lambda,T)$ is the Planck
function, $T_{\rm w}$ and $T_{\rm c}$ are the warm and cold dust temperatures
and $Q_{\rm em}$ is
the wavelength-dependent emissivity of the grains which varies as $\lambda^{- \beta}$
over the wavelength range considered. We further assume that $\beta$ has a fixed
value of 2 which is the typical value found for galaxies (see e.g. the
discussion in Dunne \& Eales 2001). We thus have four free parameters; the two 
normalizations and the two temperatures. 

Five of the galaxies have enough data points to constrain this model by
minimizing $\chi^2$. These fits are presented in Table~\ref{table:modres} and
Fig.~\ref{fig:gmodels}. In Fig.~\ref{fig:gmodels} the confidence limits for the
fitted temperatures can be estimated roughly by the reader by projecting the
$\chi^2$ contours onto the axes. Note that these estimates give the confidence
limits for our adopted model and are not absolute. The reduced $\chi^2$ values
and probabilities show that all fits are acceptable (values of $Q>0.001$ are
often assumed to indicate adequate fits to real data). For the remaining ten
galaxies, which all have four data points, the fits are formally unconstrained;
we simply give the minimum $\chi^2$ in Table~\ref{table:modres} and show these
fits in Fig.~\ref{fig:omodels}. Inspection of Table~\ref{table:modres} shows
that constrained and unconstrained fits return temperatures and normalization
ratios with similar ranges. We thus use the full sample in subsequent analysis.

First, we calculate the dust mass of each component in the standard manner using,
\begin{equation}
M_{\rm d} = S(\lambda)D^2/[k_{\rm d}(\lambda)B(\lambda ,T)]
\end{equation}
where $D$ is the distance to the galaxy and $k_{\rm d}$ is the mass absorption
coefficient for which we assume a value $0.077$~m$^2$~kg$^{-1}$ at 850~$\mu$m
(Draine \& Lee 1984). Note that the ratio of the fitted normalizations gives
the ratio of the masses in the warm and cold dust components. We then
calculate the ratio of luminosity in the cold component to that in the warm by
integrating under the fitted curves. The `far-infrared' luminosity ($L_{\rm FIR}$) 
is calculated similarly by summing the luminosity in both components. 
All of these quantities are given in Table~\ref{table:modres}. 
 
\begin{table*}
\begin{minipage}{4.5in}
\caption{Model fitting results.}
\label{table:modres}
\small
\begin{tabular}{lcccccccc}\hline
(1) & (2) & (3) & (4) & (5) & (6) & (7) & (8) & (9) \\
Source & $T_{\rm w}$ & $T_{\rm c}$ & $N_{\rm c}/N_{\rm w}$ &
$M_{\rm d}$ & $L_{\rm c}/L_{\rm w}$ & $L_{\rm FIR}$ & $\chi^2/\nu$ & $Q$ \\
& (K) & (K) & & (Log\ ${\rm M}_{\odot}$) & &
(Log\ ${\rm L}_{\odot}$) & &\\ \hline
NGC\ 660& $32$ & $10$ & $28$ & $8.34$ & $0.02$ & $10.69$ & $1.5$ & $0.23$\\
NGC\ 3079& $29$ & $13$  & $7$ & $8.11$ & $0.06$ & $10.84$ & $1.0$ & $0.40$\\
NGC\ 4402& $25$ & $13$ & $6$ & $7.55$ & $0.14$ & $9.88$ & $0.2$ & $0.70$ \\
NGC\ 4631& $38$ & $17$ & $99$ & $8.12$ & $0.75$ & $10.62$ & $5.3$ & $1.2\times10^{-3}$\\
NGC\ 5907& $28$ & $16$ & $48$ & $7.94$ & $1.76$ & $ 10.15$ & $0.2$ & $0.66$ \\
NGC\ 157& $28$ & $9$ & $22$ & $8.67$ & $0.03$ & $10.75$ & \ldots & \ldots \\
NGC\ 1808& $32$ & $11$ & $13$ & $8.03$ & $0.02$ & $ 10.70$ & \ldots & \ldots\\
NGC\ 2903& $28$ & $5$ & $46$ & $8.35$  & $<0.01$ & $10.11$ & \ldots & \ldots\\
NGC\ 3310& $33$ & $8$ & $300$ & $9.04$ & $0.06$ & $10.47$ & \ldots & \ldots\\
NGC\ 3368& $27$ & $12$ & $ 8$ & $7.48$ & $0.06$ & $9.89$ & \ldots & \ldots\\
NGC\ 3628& $29$ & $10$ & $ 37$ & $7.72$ & $0.05$ & $9.74$& \ldots & \ldots \\
NGC\ 4303& $29$ & $9$  & $9$ & $7.97$ & $0.01$ & $10.55$ & \ldots & \ldots\\
NGC\ 4388& $31$ & $14$ & $17$ & $7.37$ & $0.18$ & $9.86$ & \ldots & \ldots\\
NGC\ 4414& $29$ & $12$ & $11$ & $7.78$ & $0.06$ & $10.20$ & \ldots & \ldots\\
NGC\ 4501& $25$ & $10$ & $5$ & $8.00$ & $0.21$ & $10.41$ & \ldots & \ldots\\ \hline
\end{tabular}
(1) Source name. (2) Cold dust temperature. (3) Warm dust temperature.
(4) Ratio of cold-to-warm dust. (5) Total dust mass. (6) Ratio of luminosity in the cold
    component with luminosity in the hot component. Luminosities are calculated by 
integrating under the fitted curves. (7)
    Far-infrared luminosity calculated by integrating under both fitted
    curves. (8) Minimum reduced $\chi^2$. (9) $Q$ gives the probability that
the observed reduced chi-square will exceed the value $\chi^2$ by chance even for a
correct model.
The first five rows show results for galaxy spectra with more than 4 data
    points whereas the remainder have only 4 data points and the fits are unconstrained.
\end{minipage}
\end{table*}

\section{Discussion}

\subsection{Gas and dust}

\begin{figure*}
\setlength{\unitlength}{1in}
\begin{picture}(6.0,2.5)
\includegraphics{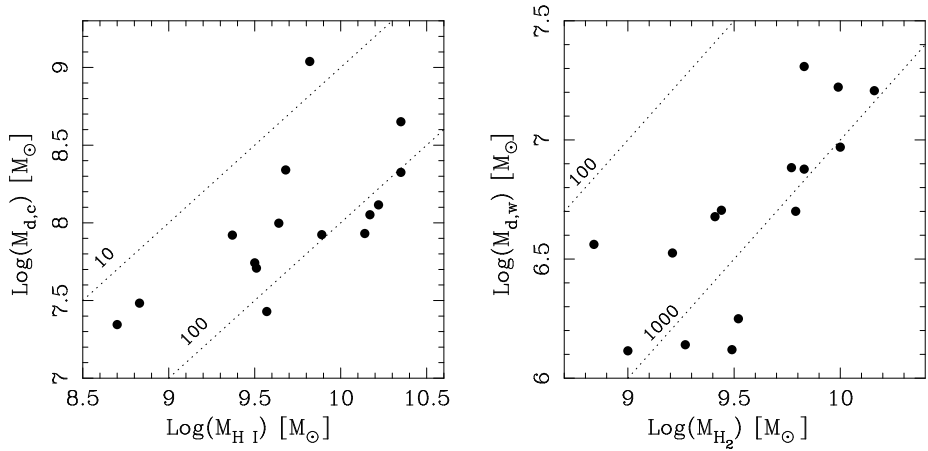}
\end{picture}
\caption[dum]{Correlations between (left) mass of atomic hydrogen and mass
of dust in the cold component, and (right) mass of molecular
hydrogen and mass of dust in the warm component . Both
correlations are significant at the $>3$-$\sigma$ level (see text). The dotted
lines on each panel are gas-to-dust mass ratios of the indicated level.}
\label{fig:gasdust}
\end{figure*}

\begin{table*}
\begin{minipage}{123mm}
\caption{Gas masses from the literature, calculated gas-to-dust mass ratios and
star-formation efficiencies.}
\label{table:gaslit}
\begin{tabular}{lccccccc}\hline
(1)&(2)&(3)&(4)&(5)&(6)&(7)&(8)\\
Source & $M_{\rm H_I}$ & $M_{\rm{H_2}}$ & $M_{\rm H_I}/M_{\rm d,c}$ & $M_{\rm
H_2}/M_{\rm d,w}$ & $M_{\rm g}/M_{\rm d}$ & $L_{\rm w}/M_{\rm{H_2}}$ & 
$L_{\rm FIR}/M_{\rm g}$\\ 
&($\rm{Log~M_{\odot}}$)&($\rm{Log~M_{\odot}}$)
& & & & (${\rm L}_{\odot}\,{\rm M}_{\odot}^{-1}$) & (${\rm L}_{\odot}\,{\rm
M}_{\odot}^{-1}$ )\\\hline
NGC 157&10.35$^a$&        9.83& 50 & 332 & 63 & 8.1 & 1.9 \\
NGC 660&10.09$^b$&       10.08& 106 & 896 & 110 & 7.1 & 1.7 \\
NGC 1808&9.64$^c$&   (i)  9.77& 44 & 769 & 95 & 8.3 & 4.9 \\
NGC 2903&9.68$^d$&        9.41& 22 & 539 & 33 & 5.0 & 1.8 \\
NGC 3079&10.17&          10.16& 131 & 898 & 229 & 4.5 & 2.4\\
NGC 3310&9.82$^e$&        8.84& 6 & 190 & 7 &40.2 & 4.0 \\
NGC 3368&  9.57$^f$&      9.21& 138 & 483 & 174 & 4.5 & 1.5  \\
NGC 3628&  9.51&   (ii)   9.27& 63 & 1348 & 98 & 2.8 & 1.1 \\
NGC 4303& 9.89$^g$&      10.00& 92 & 1072 & 191 & 3.5 & 2.0 \\
NGC 4388& 8.70$^g$&       9.00& 23 & 768 & 65 & 6.1 & 4.8 \\
NGC 4402&  8.83$^h$&      9.44& 22 & 543 & 98 & 2.4 & 2.2  \\
NGC 4414& 9.50$^i$&       9.79& 57 & 1228 & 155 & 2.4 & 1.7\\
NGC 4501&  9.37$^g$&      9.99& 28 & 586 & 120 & 2.2 & 2.1  \\
NGC 4631& 10.22$^j$&      9.49& 127 & 2344 & 148 & 7.7 & 2.1  \\
NGC 5907& 10.14$^k$&      9.52& 162 & 1863 & 195 & 3.6 & 0.8  \\
&&&&&\\
\multicolumn{3}{r}{Mean} & $71\pm49$ & $924\pm562$ & $119\pm61$ & $7.2\pm9.1$ &
$2.3\pm1.2$\\ \hline
\end{tabular}

(1) Source name. (2) Mass of atomic hydrogen. (3) Mass of molecular
hydrogen. (4) ratio between masses in atomic hydrogen and cold dust. (5) ratio
between masses in molecular hydrogen and warm dust (6) Total gas-to-dust mass
ratio. (7) star-formation efficiency calculated from the luminosity in the warm
component and the mass in molecular hydrogen. (8) global star-formation efficiency. 
References:
 a -- Ryder et~al. (1998);
 b -- van Driel et~al. (1995);
 c -- Dahlem, Ehle \& Ryder (2001);
 d -- Wevers, van der Kruit \& Allen (1986);
 e -- Mulder, van Driel \& Braine (1995);
 f -- Devereux \& Young (1990; reference within to Warmels 1986);
 g -- Cayatte et~al. (1990);
 h -- Giovanelli \& Haynes (1983);
 i -- Braine, Combes \& van Driel (1993);
 j -- Rand (1994);
 k -- Dumke et~al. (1997; reference within to Huchtmeier \& Richter 1989);
 All H$_2$ masses are from Young et~al. (1995) {\it except}
    (i)  Mass taken from Haynes, Giovanelli \& Roberts (1979; scaled for distance).
    (ii)  Mass taken from Dahlem et~al. (1990; scaled for distance and X$\rm{_{CO}}$).
\end{minipage}
\end{table*}

\begin{figure}
\setlength{\unitlength}{1in}
\begin{picture}(3.0,8.3)
\includegraphics{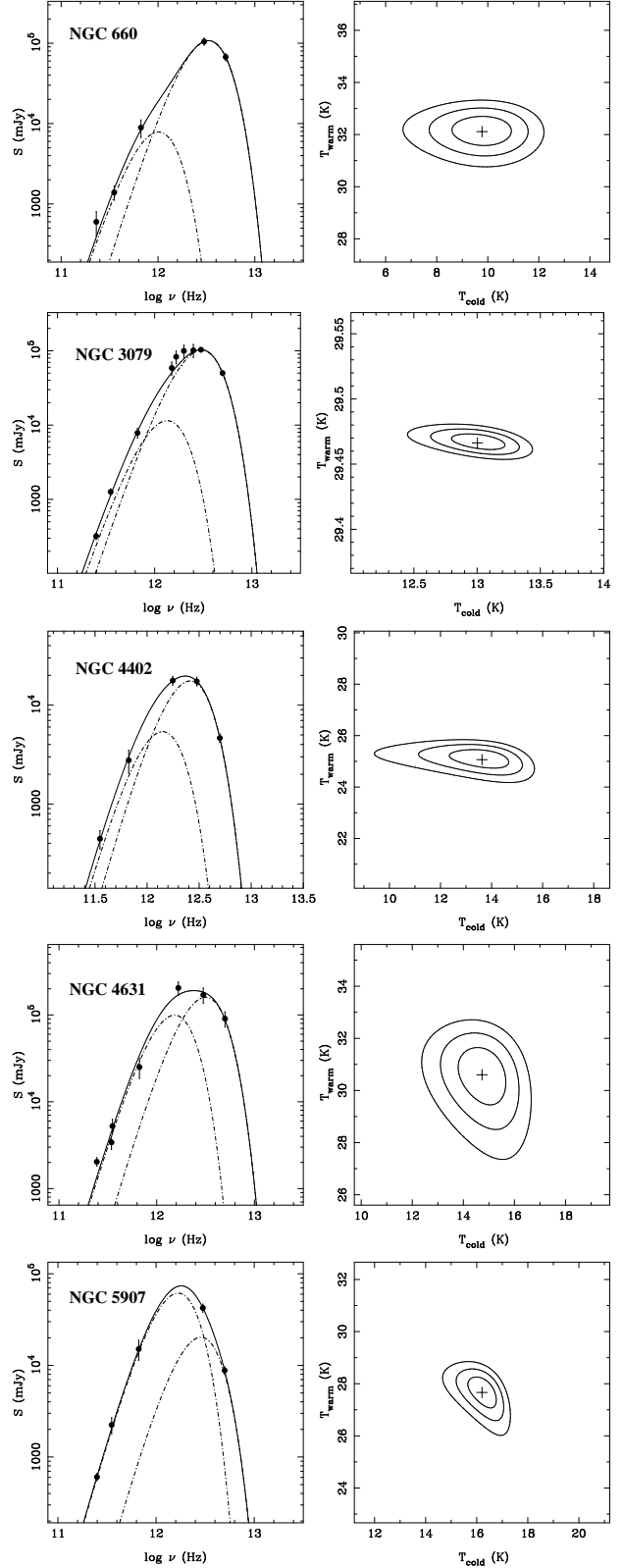}
\end{picture}
\caption[dum]{Model fits for those galaxies with more than 4 data points. On the
left-hand panels the dot-dashed lines show the warm and cold component
optically-thin greybody fits with a fixed emissivity index, $\beta=2$. The
solid line is the sum of these two components. The right-hand panels show
contours of $\chi^2$ at 1-, 2- and 3-$\sigma$ confidence intervals for the fitted
temperatures calculated while keeping the fitted normalizations fixed. The
cross on each right-hand panel marks the position of minimum $\chi^2$.
}
\label{fig:gmodels}
\end{figure}

\begin{figure}
\setlength{\unitlength}{1in}
\begin{picture}(3.0,8.3)
\includegraphics{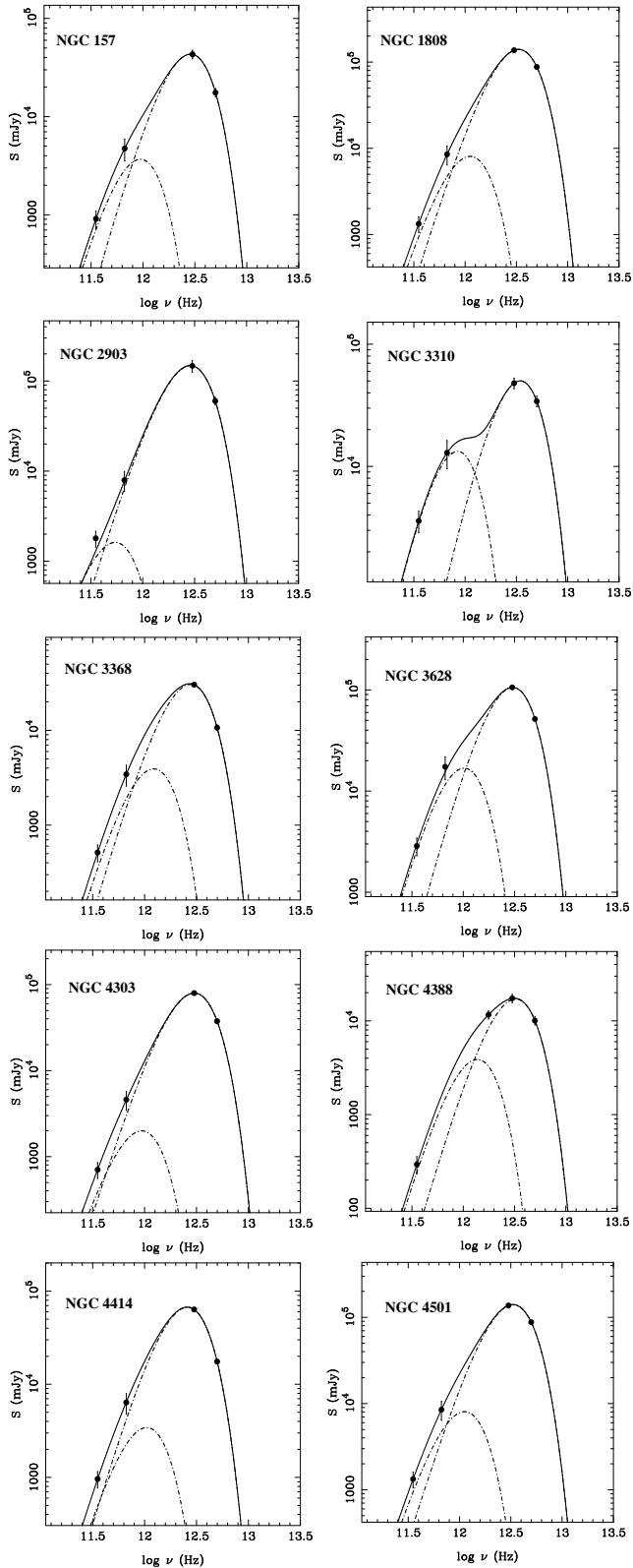}
\end{picture}
\caption[dum]{Minimum $\chi^2$ fits for those galaxies with only 4 data points.
The dot-dashed lines show the warm and cold component
optically thin greybody fits with a fixed emissivity index, $\beta=2$. The
solid line is the sum of these two components. 
}
\label{fig:omodels}
\end{figure}

\begin{figure*}
\setlength{\unitlength}{1in}
\begin{picture}(7.0,2.2)
\includegraphics{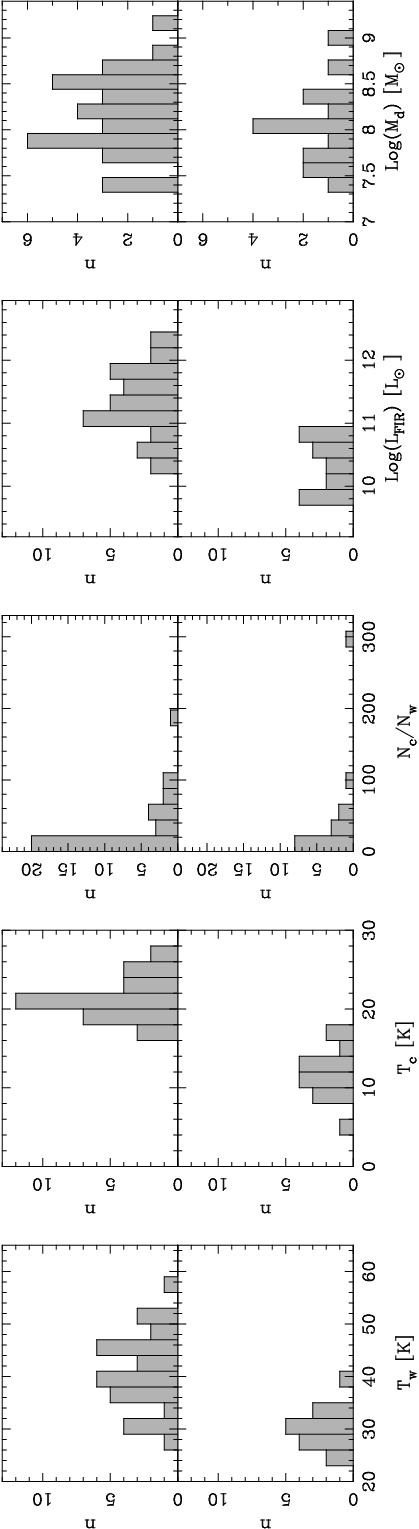}
\end{picture}
\caption[dum]{Histograms of quantities derived from the model fits showing (from left
to right) the distributions of warm temperature, cold temperature, ratio of
mass in the cold to mass in the warm component, logarithm of far-infrared
luminosity and logarithm of the dust mass. The top panels show data from the
SLUGS survey (Dunne \& Eales 2001) while the bottom panels show data from this
work.  
}
\label{fig:hists}
\end{figure*}

All of our galaxies have been mapped in the CO(1-0) line, a tracer of molecular
hydrogen, and the 21~cm line of atomic hydrogen. The gas masses computed from
these data are now used together with the dust masses from the model fits to
search for correlations between these quantities, and to quantify the
gas-to-dust mass ratios. Global CO(1-0) fluxes were taken from the Five
Colleges Radio Astronomy Observatory Extragalactic CO Survey (Young
et~al. 1995) and converted to molecular gas masses ($M_{\rm H_2}$) using a
conversion factor $X_{\rm CO} = 2.8 \times
10^{20}$~H$_2$~mol~cm$^{-2}$\,(K~km~s$^{-1}$)$^{-1}$. Atomic hydrogen masses
($M_{\rm H_I}$) were taken from the literature. These quantities are given in
Table~\ref{table:gaslit} along with references to the literature sources. Also
given are ratios between the H~{\sc i} mass and cold dust mass ($M_{\rm d,c}$),
the H$_2$ mass and warm dust mass ($M_{\rm d,w}$) and the total gas-to-dust
mass ratio ($M_{\rm g}/M_{\rm d}$) where $M_{\rm g}=M_{\rm H_2}+M_{\rm H_I}$.

In Section~\ref{section:mod} we raised the idea that the cold dust might trace
the atomic hydrogen and the warm dust the molecular hydrogen. The dust masses
of these components derived from our model fits are compared with the relevant
gas mass in Fig.~\ref{fig:gasdust}. A Spearman rank-order test returns a
correlation coefficient, $r=0.78$ and probability that the correlation is real,
$P>0.999$ for the former case and $r=0.79$, $P>0.999$ for the latter. Both
correlations are significant at $>3$-$\sigma$ confidence.  However, no
correlation is found between either cold dust mass and molecular hydrogen mass
($r=0.21$, $P=0.55$) or between warm dust mass and atomic hydrogen mass
($r=0.26$, $P=0.66$). These results lend weight to our assertion and show that
the model fits are returning physically meaningful quantities.  An obvious next
step would be to investigate the spatial correspondence between dust and gas
emission, and to search for variations in submillimetre spectral index as a
function of position within the galaxies.  However, since such detailed
decompositions do not readily fit the scope of this work we defer them to a
future paper.

The mean total gas-to-dust mass ratio of $119\pm61$ is a factor $\sim10$ below
the mean value derived by DY90 ($1080\pm70$) from {\em IRAS\/} data alone and
entirely equivalent to the mean value within our own galaxy. For the latter,
Sodroski et al (1997) report gas-to-dust mass ratios of $\sim70-270$ for the
H~{\sc i} component and $\sim140-400$ for the H$_2$ component, both showing a
systematic decline towards the Galactic centre (we have applied a correction
factor to take account of the different mass absorption coefficient used in
that study compared to ours). These ratios are consistent with the mean values
from this study given in Table~\ref{table:gaslit}.  The fact that in a sample
of similar spiral galaxies to our own we obtain a very similar global ratio
suggests we have now identified essentially all of the dust in these sources.

Although the scatter is large, there is evidence that the gas-to-dust mass
ratios in the warm (molecular) components are systematically higher than those
in the cold (atomic) components (Table~\ref{table:gaslit} and
Fig.~\ref{fig:gasdust}). This result is, however, contrary to that expected
since the gas-to-dust mass ratio in galaxies is known to decrease with
increasing metallicity (e.g. Whittet 1992). One possibility might be that
$X_{\rm CO}$ varies as a function of environment. In fact, for NGC~3079, Braine
et al. (1997) calculate a value for the core region that is roughly one order
of magnitude below that derived for the disk. However, they also find that the
molecular gas mass of NGC~3079 is dominated by the disk component so while this
effect does not lead to an order of magnitude reduction in the gas-to-dust mass
ratio, it would reduce it to some extent, and possibly by a more substantial
amount in other galaxies. On the other hand, the gas-to-dust mass ratios might
be raised in these active regions if dust grain mantles are destroyed while the
CO molecules are left intact.  

It has been argued that a good measure of the activity of a galaxy is its
star-formation efficiency which is simply the efficiency of converting mass
into luminosity, and is defined as $L_{\rm FIR}/M_{\rm g}$ (e.g. Chini et
al. 1986). Very active star-forming galaxies are found to have $L_{\rm
FIR}/M_{\rm g}\sim100$ ${\rm L}_{\odot}\,{\rm M}_{\odot}^{-1}$ while quiescent
systems such as spiral galaxies have $L_{\rm FIR}/M_{\rm g}\sim5$ ${\rm
L}_{\odot}\,{\rm M}_{\odot}^{-1}$ (Chini et al. 1995).  In
Table~\ref{table:gaslit} we list the star-formation efficiencies for our spiral
galaxies; we give both the global measure and the measure for the warm
molecular component. As expected the star-formation efficiencies are higher for
the latter but in only one case (NGC~3310) do they approach the typical value
found for active systems. While this could be taken as evidence that these
galaxies do not contain regions of substantial active star-formation there is
inevitably some degree of `smearing' when measuring the global properties of
these systems. A more detailed comparison of the dust emission and molecular
gas emission is required. For example, Braine et al. (1997) used millimetre
wavelength dust continuum and CO observations of similar spatial resolution to
those presented here to show that the central region of NGC~3079 has $L_{\rm
FIR}/M_{\rm g}\sim100$ ${\rm L}_{\odot}\,{\rm M}_{\odot}^{-1}$ whereas the disk
has $L_{\rm FIR}/M_{\rm g}\sim6-7$ ${\rm L}_{\odot}\,{\rm M}_{\odot}^{-1}$.

\subsection{Dust masses, temperatures and luminosities}

It is instructive to compare the model fitted parameters from
Section~\ref{section:mod} with those determined for other galaxies, including
our own Milky Way. We first consider results for SLUGS galaxies which were
selected from the {\em IRAS\/} Bright Galaxy Sample (Dunne \& Eales 2001). The
models used in this paper are identical to those fitted to the SLUGS galaxies,
and we have used the same dust parameters (mass absorption coefficient). The two
samples, once corrected for different assumed cosmologies, are thus directly
comparable. We plot histograms of selected quantities from
Table~\ref{table:modres} in Fig.~\ref{fig:hists}. It is apparent that our
heterogeneously-selected sample has a lower mean far-infrared luminosity
$\{$Log[$\overline{L}_{\rm FIR}\ ({\rm L}_{\odot})]=10.3\pm0.4\}$ than the
SLUGS sample $\{$Log[$\overline{L}_{\rm FIR}\ ({\rm
L}_{\odot})]=11.0\pm0.5\}$. This difference also appears to hold for the
distributions of warm dust temperature ($\overline{T}_{\rm w}=30\pm3$~K
cf. $\overline{T}_{\rm w}=41\pm7$~K for SLUGS) and cold dust temperature
($\overline{T}_{\rm c}=11\pm3$~K cf.  $\overline{T}_{\rm c}=21\pm3$~K for
SLUGS). The far-infrared-selected galaxies have on average higher dust
temperatures for both components. The distributions of normalization ratio and
dust mass, however, look similar. These results are confirmed with the
application of two-sample Kolmogorov-Smirnov (K-S) tests
(Table~\ref{table:ks}).

What is the physical origin of this difference in dust temperature?  The
equilibrium temperature of a dust grain is a function of both the grain
properties and the energy density in the ISRF. While it is possible that the
properties of the dust grains differ in the warm and cold components the fact
that the lower luminosity galaxies have lower dust temperatures argues for a
connection between star-formation activity and dust temperature. Global
star-formation efficiencies can be calculated for 14 of the SLUGS galaxies. As
expected, they are typically higher than those found for our sample
(Table~\ref{table:gaslit}) with a range $2-32$ ${\rm L}_{\odot}\,{\rm
M}_{\odot}^{-1}$ and a mean of $9\pm9$ ${\rm L}_{\odot}\,{\rm M}_{\odot}^{-1}$.
Since the temperatures of both components appear to vary in concert the nature
of the star-formation process must be such that the heating is not confined
solely to dust in molecular clouds. Some fraction of energy emitted by OB stars
must escape the molecular cores and heat dust in less dense regions. Thus an
increase in star-formation activity would have a wider impact than naively
expected from the simple two component model.

Finally in this section, we compare our results with those obtained for the
Milky Way. The total far-infrared luminosity is $7.4\times 10^9~{\rm
L}_{\odot}$ of which $\sim60$ per cent is from dust associated with the H~{\sc
i} component (Sodroski et al. 1997). Inspection of the model fits and the
$L_{\rm c}/L_{\rm w}$ values from Table~\ref{table:gaslit} shows that only
NGC~5907 has this property. The far-infrared luminosity of the other galaxies
appears to be dominated by the warmer dust component that we associate
predominantly with H$_2$. Dust temperatures for the Milky way are presented by
Reach et al. (1995) who find three characteristic ranges of $16-21$, $10-13$
and $4-7$~K. Thus in the context of our sample of spiral galaxies the Milky Way
is less luminous and contains dust radiating at lower temperatures.
    
\begin{table}
\begin{center}
\begin{minipage}{40mm}
\caption{K-S test results.}
\label{table:ks}
\begin{tabular}{lcc}\hline
(1) & (2) & (3) \\
Samples & KS & probability \\ \hline
$T_{\rm w}$ & 0.75 & $>0.999$ \\    
$T_{\rm c}$ & 0.97 & 1.00 \\    
$N_{\rm c}/N_{\rm w}$ & 0.18 & 0.16 \\    
$L_{\rm FIR}$ & 0.78 & $>0.999$ \\    
$M_{\rm d}$ & 0.26 & 0.59 \\\hline
\end{tabular}   
(1) Test samples. (2) K-S statistic. (3) Probability that datasets are drawn
    from different distributions. 
\end{minipage}
\end{center}
\end{table}

\subsection{Effect of environment}

Because we have a mixture of field and Virgo cluster spirals in our sample we
can try to make a comparison of the gas to dust ratios in the different
environments. It is well known that some Virgo spirals are deficient in H~{\sc
i}, presumably due to stripping, however H$_2$ is not depleted (Kenney \& Young
1986, 1988a,b, 1989; Stark et al. 1986). Leggett, Brand \& Mountain (1987)
claimed that there was no difference in the {\em IRAS\/} emission from field
and Virgo spirals, consistent with the idea that {\em IRAS\/} traced warm dust
associated with well-bound molecular clouds. However, Doyon \& Joseph (1989)
found evidence that H~{\sc i} deficient galaxies in the Virgo cluster have
lower 60- and 100-$\mu$m flux densities and cooler far-infrared colour
temperatures than those with normal H~{\sc i} content. They conclude that for a
typical spiral in the core of the cluster, at least half the diffuse dust has
been stripped.  If true then then we might expect to see a difference at longer
wavelengths.  Specifically, if the cold dust that is mixed with the H~{\sc i}
is also stripped from the Virgo galaxies then this difference would be
reflected in the $N_{\rm c}/N_{\rm w}$ ratios determined in the model fits
since these give the mass of dust in the cold component relative to that in the
warm component. The values found for the Virgo galaxies are indeed all
relatively small (9, 17, 6 and 5) with a mean value of just $9\pm5$. For
comparison, the non-Virgo members have a mean value of $56\pm81$ or $31\pm26$
excluding NGC~3310. However, the samples are not formally distinct; a
two-sample K-S test returns a KS statistic of 0.64 with a probability of 0.89
(1.6 $\sigma$) that the two datasets are drawn from different
distributions. Observations of much larger samples would be required to
investigate the effect of environment any further.

\section{Conclusions}

We have presented high-quality submillimetre images of 14 spiral
galaxies. Simple two-component model fits to the global SEDs show that:

\noindent
(1) The dust is typically radiating at temperatures of $10-20$~K and $25-40$~K
    and in most cases the warm component dominates the total luminosity.

\noindent
(2) The dust mass in the warm component correlates with the mass in molecular
    hydrogen while the dust mass in the cold component correlates with the mass
    in atomic hydrogen. The opposite correlations are not significant. This
    result suggests that the simple two-component models provide a good
    approximation to the real physical nature of these galaxies.

\noindent
(3) The mean gas-to-dust mass ratio of $120\pm60$ is entirely consistent with
    that reported for the Milky Way. There is evidence that the ratio is
    higher in the warm, molecular component.

\noindent
(4) Comparing our model results with those found for the SLUGS survey, we find
    that our galaxies have, on average, similar dust masses but lower
    far-infrared luminosities and lower dust temperatures for both the warm and cold
    components. We suggest that the lower dust temperatures found for our
    galaxies can be linked to reduced star-formation activity and
    thus less intense ISRFs than those present in the SLUGS galaxies. \\

The next step will be to analyse the emission from the individual galaxies in
more detail. In particular to investigate both the correspondence of dust
and gas emission and to search for differences in submillimetre spectral index
as a function of galactocentric radius.   

\section*{ACKNOWLEDGMENTS}

MA acknowledges the support of a PPARC studentship.  We thank Loretta Dunne for
insightful comments on the pre-submitted manuscript, Ashley James for helpful
discussions on dust properties and the referee for constructive
criticism.  Guest User, Canadian Astronomy Data Centre, which is operated by
the Dominion Astrophysical Observatory for the National Research Council of
Canada's Herzberg Institute of Astrophysics. This research has made use of the
NASA/IPAC Extragalactic Database (NED) which is operated by the Jet Propulsion
Laboratory, California Institute of Technology, under contract with the
National Aeronautics and Space Administration. The DSS images presented in this
paper were obtained from the Multimission Archive at the Space Telescope
Science Institute (MAST). STScI is operated by the Association of Universities
for Research in Astronomy, Inc., under NASA contract NAS5-26555. Support for
MAST for non-HST data is provided by the NASA Office of Space Science via grant
NAG5-7584 and by other grants and contracts.

\bsp
\label{lastpage}
\end{document}